\documentclass[11pt]{article}
\usepackage{amsthm,amsmath,latexsym,multibox,amssymb,amsfonts,amsbsy}
\usepackage{lscape,fancyhdr,array,stmaryrd,euscript,wrapfig}
\usepackage{cite,wrapfig,color}
\usepackage{graphicx}

\newcommand{\bep}{\begin{picture}}
\newcommand{\eep}{\end{picture}}

\newcolumntype{x}[1]{%
>{\centering\hspace{0pt}}m{#1}}%
\newcolumntype{w}[1]{%
>{\raggedright\hspace{0pt}}m{#1}}%
\newcolumntype{z}[1]{%
>{\raggedleft\hspace{0pt}}m{#1}}%

{\vspace{3mm} }

\def\al{\alpha}

\def\*{\star}

\def\E2{\mathbf{E}}

\makeatletter

\newcommand{\Rmnum}[1]{\expandafter\@slowromancap\romannumeral #1@}
\makeatother

\newcommand{\be}{\begin{equation}}
\newcommand{\ee}{\end{equation}}
\newcommand{\bee}{\begin{eqnarray}}
\newcommand{\beee}{\begin{array}}
\newcommand{\eee}{\end{eqnarray}}
\newcommand{\eeee}{\end{array}}
\newcommand{\gc}{\chi}

\newcommand{\gad}{\dot{\alpha}}
\newcommand{\gbd}{\dot{\beta}}
\newcommand{\gdd}{\dot{\gamma}}
\newcommand{\ga}{\alpha}
\newcommand{\gb}{\beta}

\newcommand{\gd}{\delta}
\newcommand{\gl}{\lambda}

\newcommand{\gep}{\epsilon}

\newcommand{\gO}{\Omega}

\newcommand{\dal}{\dot \alpha}
\newcommand{\dgb}{\dot \beta}

\newcommand{\nn}{\nonumber}

\newcommand{\p}{\partial}
\newcommand{\pl}{\partial}

\newcommand{\ff}{\frac}
\newcommand{\fud}[2]{{}^{#1}{}_{#2}\,}
\newcommand{\fdu}[2]{{}_{#1}{}^{#2}\,}

\newcommand{\rmx}{{\mathrm{x}}}
\newcommand{\mcG}{{\mathcal{G}}}

\newcommand{\QstrA}{\parbox{63pt}{
\bep(50,25)(-10,-4)
\put(0,0){\circle*{4}}\put(-6,5){$\scriptstyle i\!-\!1$}
\put(2,1){\vector(2,1){16}}
\put(20,10){\circle*{4}}\put(19,15){$\scriptstyle i$}
\put(22,9){\vector(2,-1){16}}
\put(40,0){\circle*{4}}\put(39,5){$\scriptstyle i\!+\!1$}
\eep}
}

\newcommand{\QstrB}{\parbox{63pt}{
\bep(50,25)(-10,-4)
\put(0,0){\circle*{4}}\put(-6,5){$\scriptstyle i\!+\!1$}
\put(18,9){\vector(-2,-1){16}}
\put(20,10){\circle*{4}}\put(19,15){$\scriptstyle i$}
\put(38,1){\vector(-2,1){16}}
\put(40,0){\circle*{4}}\put(39,5){$\scriptstyle i\!-\!1$}
\eep}
}

\newcommand{\PstrA}{\parbox{29pt}{
\bep(38,12)(-8,-3)
\put(0,0){\circle*{4}}\put(-5,5){$\scriptstyle i\!-\!1$}
\put(3,0){\vector(1,0){10}}
\put(15,0){\circle*{4}}\put(14,5){$\scriptstyle i$}
\eep}
}

\newcommand{\PstrB}{\parbox{29pt}{
\bep(38,12)(-5,-3)
\put(0,0){\circle*{4}}\put(-1,5){$\scriptstyle i$}
\put(12,0){\vector(-1,0){10}}
\put(15,0){\circle*{4}}\put(10,5){$\scriptstyle i\!-\!1$}
\eep}
}

\newcommand{\NgonA}{\parbox{50pt}{
\bep(50,50)(-3,-32)
\put(0,0){\circle*{4}}\put(-1,5){$\scriptstyle n$}
\put(2,1){\vector(2,1){16}}
\put(20,10){\circle*{4}}\put(19,15){$\scriptstyle 1$}
\put(22,9){\vector(2,-1){16}}
\put(40,0){\circle*{4}}\put(39,5){$\scriptstyle 2$}
\put(40,-1){\vector(0,-1){17}}
\put(40,-20){\circle*{4}}\put(44,-23){$\scriptstyle 3$}
\put(0,-19){\vector(0,1){17}}
\put(0,-20){\circle*{4}}
\put(18,-29){\vector(-2,1){16}}
\put(20,-30){\circle*{4}}
\put(38,-21){\vector(-2,-1){16}}
\eep}
}

\newcommand{\NgonB}{\parbox{50pt}{
\bep(50,50)(-8,-32)
\put(0,0){\circle*{4}}\put(-1,5){$\scriptstyle 2$}
\put(18,9){\vector(-2,-1){16}}
\put(20,10){\circle*{4}}\put(19,15){$\scriptstyle 1$}
\put(38,1){\vector(-2,1){16}}
\put(40,0){\circle*{4}}\put(39,5){$\scriptstyle n$}
\put(40,-19){\vector(0,1){17}}
\put(40,-20){\circle*{4}}\put(-7,-23){$\scriptstyle 3$}
\put(0,-1){\vector(0,-1){17}}
\put(0,-20){\circle*{4}}
\put(2,-21){\vector(2,-1){16}}
\put(20,-30){\circle*{4}}
\put(22,-29){\vector(2,1){16}}
\eep}
}

\newcommand{\TwistorialWitten}{
\ifx\JPicScale\undefined\def\JPicScale{0.8}\fi
\unitlength \JPicScale mm
\begin{picture}(40.26,40.76)(-5,40)
\linethickness{0.2mm}
\put(18,53){$X$}\put(8,84){$x_1$}\put(40,70){$x_2$}\put(-10,60){$x_n$}
\put(40.25,61){\line(0,1){0.5}}
\put(19,61){\circle*{5}}
\multiput(40.24,62)(0.01,-0.5){1}{\line(0,-1){0.5}}
\multiput(40.22,62.49)(0.02,-0.5){1}{\line(0,-1){0.5}}
\multiput(40.19,62.99)(0.03,-0.5){1}{\line(0,-1){0.5}}
\multiput(40.14,63.48)(0.05,-0.49){1}{\line(0,-1){0.49}}
\multiput(40.08,63.97)(0.06,-0.49){1}{\line(0,-1){0.49}}
\multiput(40.01,64.47)(0.07,-0.49){1}{\line(0,-1){0.49}}
\multiput(39.93,64.96)(0.08,-0.49){1}{\line(0,-1){0.49}}
\multiput(39.84,65.45)(0.09,-0.49){1}{\line(0,-1){0.49}}
\multiput(39.74,65.93)(0.1,-0.49){1}{\line(0,-1){0.49}}
\multiput(39.63,66.42)(0.11,-0.48){1}{\line(0,-1){0.48}}
\multiput(39.5,66.9)(0.12,-0.48){1}{\line(0,-1){0.48}}
\multiput(39.37,67.37)(0.14,-0.48){1}{\line(0,-1){0.48}}
\multiput(39.22,67.85)(0.15,-0.47){1}{\line(0,-1){0.47}}
\multiput(39.06,68.32)(0.16,-0.47){1}{\line(0,-1){0.47}}
\multiput(38.89,68.79)(0.17,-0.47){1}{\line(0,-1){0.47}}
\multiput(38.71,69.25)(0.18,-0.46){1}{\line(0,-1){0.46}}
\multiput(38.52,69.71)(0.1,-0.23){2}{\line(0,-1){0.23}}
\multiput(38.32,70.16)(0.1,-0.23){2}{\line(0,-1){0.23}}
\multiput(38.11,70.61)(0.11,-0.22){2}{\line(0,-1){0.22}}
\multiput(37.89,71.06)(0.11,-0.22){2}{\line(0,-1){0.22}}
\multiput(37.66,71.5)(0.12,-0.22){2}{\line(0,-1){0.22}}
\multiput(37.42,71.93)(0.12,-0.22){2}{\line(0,-1){0.22}}
\multiput(37.16,72.36)(0.13,-0.21){2}{\line(0,-1){0.21}}
\multiput(36.9,72.78)(0.13,-0.21){2}{\line(0,-1){0.21}}
\multiput(36.63,73.2)(0.14,-0.21){2}{\line(0,-1){0.21}}
\multiput(36.35,73.61)(0.14,-0.2){2}{\line(0,-1){0.2}}
\multiput(36.06,74.01)(0.15,-0.2){2}{\line(0,-1){0.2}}
\multiput(35.76,74.41)(0.15,-0.2){2}{\line(0,-1){0.2}}
\multiput(35.45,74.8)(0.1,-0.13){3}{\line(0,-1){0.13}}
\multiput(35.14,75.18)(0.11,-0.13){3}{\line(0,-1){0.13}}
\multiput(34.81,75.55)(0.11,-0.12){3}{\line(0,-1){0.12}}
\multiput(34.47,75.92)(0.11,-0.12){3}{\line(0,-1){0.12}}
\multiput(34.13,76.28)(0.11,-0.12){3}{\line(0,-1){0.12}}
\multiput(33.78,76.63)(0.12,-0.12){3}{\line(1,0){0.12}}
\multiput(33.42,76.97)(0.12,-0.11){3}{\line(1,0){0.12}}
\multiput(33.05,77.31)(0.12,-0.11){3}{\line(1,0){0.12}}
\multiput(32.68,77.64)(0.12,-0.11){3}{\line(1,0){0.12}}
\multiput(32.3,77.95)(0.13,-0.11){3}{\line(1,0){0.13}}
\multiput(31.91,78.26)(0.13,-0.1){3}{\line(1,0){0.13}}
\multiput(31.51,78.56)(0.2,-0.15){2}{\line(1,0){0.2}}
\multiput(31.11,78.85)(0.2,-0.15){2}{\line(1,0){0.2}}
\multiput(30.7,79.13)(0.2,-0.14){2}{\line(1,0){0.2}}
\multiput(30.28,79.4)(0.21,-0.14){2}{\line(1,0){0.21}}
\multiput(29.86,79.66)(0.21,-0.13){2}{\line(1,0){0.21}}
\multiput(29.43,79.92)(0.21,-0.13){2}{\line(1,0){0.21}}
\multiput(29,80.16)(0.22,-0.12){2}{\line(1,0){0.22}}
\multiput(28.56,80.39)(0.22,-0.12){2}{\line(1,0){0.22}}
\multiput(28.11,80.61)(0.22,-0.11){2}{\line(1,0){0.22}}
\multiput(27.66,80.82)(0.22,-0.11){2}{\line(1,0){0.22}}
\multiput(27.21,81.02)(0.23,-0.1){2}{\line(1,0){0.23}}
\multiput(26.75,81.21)(0.23,-0.1){2}{\line(1,0){0.23}}
\multiput(26.29,81.39)(0.46,-0.18){1}{\line(1,0){0.46}}
\multiput(25.82,81.56)(0.47,-0.17){1}{\line(1,0){0.47}}
\multiput(25.35,81.72)(0.47,-0.16){1}{\line(1,0){0.47}}
\multiput(24.87,81.87)(0.47,-0.15){1}{\line(1,0){0.47}}
\multiput(24.4,82)(0.48,-0.14){1}{\line(1,0){0.48}}
\multiput(23.92,82.13)(0.48,-0.12){1}{\line(1,0){0.48}}
\multiput(23.43,82.24)(0.48,-0.11){1}{\line(1,0){0.48}}
\multiput(22.95,82.34)(0.49,-0.1){1}{\line(1,0){0.49}}
\multiput(22.46,82.43)(0.49,-0.09){1}{\line(1,0){0.49}}
\multiput(21.97,82.51)(0.49,-0.08){1}{\line(1,0){0.49}}
\multiput(21.47,82.58)(0.49,-0.07){1}{\line(1,0){0.49}}
\multiput(20.98,82.64)(0.49,-0.06){1}{\line(1,0){0.49}}
\multiput(20.49,82.69)(0.49,-0.05){1}{\line(1,0){0.49}}
\multiput(19.99,82.72)(0.5,-0.03){1}{\line(1,0){0.5}}
\multiput(19.5,82.74)(0.5,-0.02){1}{\line(1,0){0.5}}
\multiput(19,82.75)(0.5,-0.01){1}{\line(1,0){0.5}}
\put(18.5,82.75){\line(1,0){0.5}}
\multiput(18,82.74)(0.5,0.01){1}{\line(1,0){0.5}}
\multiput(17.51,82.72)(0.5,0.02){1}{\line(1,0){0.5}}
\multiput(17.01,82.69)(0.5,0.03){1}{\line(1,0){0.5}}
\multiput(16.52,82.64)(0.49,0.05){1}{\line(1,0){0.49}}
\multiput(16.03,82.58)(0.49,0.06){1}{\line(1,0){0.49}}
\multiput(15.53,82.51)(0.49,0.07){1}{\line(1,0){0.49}}
\multiput(15.04,82.43)(0.49,0.08){1}{\line(1,0){0.49}}
\multiput(14.55,82.34)(0.49,0.09){1}{\line(1,0){0.49}}
\multiput(14.07,82.24)(0.49,0.1){1}{\line(1,0){0.49}}
\multiput(13.58,82.13)(0.48,0.11){1}{\line(1,0){0.48}}
\multiput(13.1,82)(0.48,0.12){1}{\line(1,0){0.48}}
\multiput(12.63,81.87)(0.48,0.14){1}{\line(1,0){0.48}}
\multiput(12.15,81.72)(0.47,0.15){1}{\line(1,0){0.47}}
\multiput(11.68,81.56)(0.47,0.16){1}{\line(1,0){0.47}}
\multiput(11.21,81.39)(0.47,0.17){1}{\line(1,0){0.47}}
\multiput(10.75,81.21)(0.46,0.18){1}{\line(1,0){0.46}}
\multiput(10.29,81.02)(0.23,0.1){2}{\line(1,0){0.23}}
\multiput(9.84,80.82)(0.23,0.1){2}{\line(1,0){0.23}}
\multiput(9.39,80.61)(0.22,0.11){2}{\line(1,0){0.22}}
\multiput(8.94,80.39)(0.22,0.11){2}{\line(1,0){0.22}}
\multiput(8.5,80.16)(0.22,0.12){2}{\line(1,0){0.22}}
\multiput(8.07,79.92)(0.22,0.12){2}{\line(1,0){0.22}}
\multiput(7.64,79.66)(0.21,0.13){2}{\line(1,0){0.21}}
\multiput(7.22,79.4)(0.21,0.13){2}{\line(1,0){0.21}}
\multiput(6.8,79.13)(0.21,0.14){2}{\line(1,0){0.21}}
\multiput(6.39,78.85)(0.2,0.14){2}{\line(1,0){0.2}}
\multiput(5.99,78.56)(0.2,0.15){2}{\line(1,0){0.2}}
\multiput(5.59,78.26)(0.2,0.15){2}{\line(1,0){0.2}}
\multiput(5.2,77.95)(0.13,0.1){3}{\line(1,0){0.13}}
\multiput(4.82,77.64)(0.13,0.11){3}{\line(1,0){0.13}}
\multiput(4.45,77.31)(0.12,0.11){3}{\line(1,0){0.12}}
\multiput(4.08,76.97)(0.12,0.11){3}{\line(1,0){0.12}}
\multiput(3.72,76.63)(0.12,0.11){3}{\line(1,0){0.12}}
\multiput(3.37,76.28)(0.12,0.12){3}{\line(1,0){0.12}}
\multiput(3.03,75.92)(0.11,0.12){3}{\line(0,1){0.12}}
\multiput(2.69,75.55)(0.11,0.12){3}{\line(0,1){0.12}}
\multiput(2.36,75.18)(0.11,0.12){3}{\line(0,1){0.12}}
\multiput(2.05,74.8)(0.11,0.13){3}{\line(0,1){0.13}}
\multiput(1.74,74.41)(0.1,0.13){3}{\line(0,1){0.13}}
\multiput(1.44,74.01)(0.15,0.2){2}{\line(0,1){0.2}}
\multiput(1.15,73.61)(0.15,0.2){2}{\line(0,1){0.2}}
\multiput(0.87,73.2)(0.14,0.2){2}{\line(0,1){0.2}}
\multiput(0.6,72.78)(0.14,0.21){2}{\line(0,1){0.21}}
\multiput(0.34,72.36)(0.13,0.21){2}{\line(0,1){0.21}}
\multiput(0.08,71.93)(0.13,0.21){2}{\line(0,1){0.21}}
\multiput(-0.16,71.5)(0.12,0.22){2}{\line(0,1){0.22}}
\multiput(-0.39,71.06)(0.12,0.22){2}{\line(0,1){0.22}}
\multiput(-0.61,70.61)(0.11,0.22){2}{\line(0,1){0.22}}
\multiput(-0.82,70.16)(0.11,0.22){2}{\line(0,1){0.22}}
\multiput(-1.02,69.71)(0.1,0.23){2}{\line(0,1){0.23}}
\multiput(-1.21,69.25)(0.1,0.23){2}{\line(0,1){0.23}}
\multiput(-1.39,68.79)(0.18,0.46){1}{\line(0,1){0.46}}
\multiput(-1.56,68.32)(0.17,0.47){1}{\line(0,1){0.47}}
\multiput(-1.72,67.85)(0.16,0.47){1}{\line(0,1){0.47}}
\multiput(-1.87,67.37)(0.15,0.47){1}{\line(0,1){0.47}}
\multiput(-2,66.9)(0.14,0.48){1}{\line(0,1){0.48}}
\multiput(-2.13,66.42)(0.12,0.48){1}{\line(0,1){0.48}}
\multiput(-2.24,65.93)(0.11,0.48){1}{\line(0,1){0.48}}
\multiput(-2.34,65.45)(0.1,0.49){1}{\line(0,1){0.49}}
\multiput(-2.43,64.96)(0.09,0.49){1}{\line(0,1){0.49}}
\multiput(-2.51,64.47)(0.08,0.49){1}{\line(0,1){0.49}}
\multiput(-2.58,63.97)(0.07,0.49){1}{\line(0,1){0.49}}
\multiput(-2.64,63.48)(0.06,0.49){1}{\line(0,1){0.49}}
\multiput(-2.69,62.99)(0.05,0.49){1}{\line(0,1){0.49}}
\multiput(-2.72,62.49)(0.03,0.5){1}{\line(0,1){0.5}}
\multiput(-2.74,62)(0.02,0.5){1}{\line(0,1){0.5}}
\multiput(-2.75,61.5)(0.01,0.5){1}{\line(0,1){0.5}}
\put(-2.75,61){\line(0,1){0.5}}
\multiput(-2.75,61)(0.01,-0.5){1}{\line(0,-1){0.5}}
\multiput(-2.74,60.5)(0.02,-0.5){1}{\line(0,-1){0.5}}
\multiput(-2.72,60.01)(0.03,-0.5){1}{\line(0,-1){0.5}}
\multiput(-2.69,59.51)(0.05,-0.49){1}{\line(0,-1){0.49}}
\multiput(-2.64,59.02)(0.06,-0.49){1}{\line(0,-1){0.49}}
\multiput(-2.58,58.53)(0.07,-0.49){1}{\line(0,-1){0.49}}
\multiput(-2.51,58.03)(0.08,-0.49){1}{\line(0,-1){0.49}}
\multiput(-2.43,57.54)(0.09,-0.49){1}{\line(0,-1){0.49}}
\multiput(-2.34,57.05)(0.1,-0.49){1}{\line(0,-1){0.49}}
\multiput(-2.24,56.57)(0.11,-0.48){1}{\line(0,-1){0.48}}
\multiput(-2.13,56.08)(0.12,-0.48){1}{\line(0,-1){0.48}}
\multiput(-2,55.6)(0.14,-0.48){1}{\line(0,-1){0.48}}
\multiput(-1.87,55.13)(0.15,-0.47){1}{\line(0,-1){0.47}}
\multiput(-1.72,54.65)(0.16,-0.47){1}{\line(0,-1){0.47}}
\multiput(-1.56,54.18)(0.17,-0.47){1}{\line(0,-1){0.47}}
\multiput(-1.39,53.71)(0.18,-0.46){1}{\line(0,-1){0.46}}
\multiput(-1.21,53.25)(0.1,-0.23){2}{\line(0,-1){0.23}}
\multiput(-1.02,52.79)(0.1,-0.23){2}{\line(0,-1){0.23}}
\multiput(-0.82,52.34)(0.11,-0.22){2}{\line(0,-1){0.22}}
\multiput(-0.61,51.89)(0.11,-0.22){2}{\line(0,-1){0.22}}
\multiput(-0.39,51.44)(0.12,-0.22){2}{\line(0,-1){0.22}}
\multiput(-0.16,51)(0.12,-0.22){2}{\line(0,-1){0.22}}
\multiput(0.08,50.57)(0.13,-0.21){2}{\line(0,-1){0.21}}
\multiput(0.34,50.14)(0.13,-0.21){2}{\line(0,-1){0.21}}
\multiput(0.6,49.72)(0.14,-0.21){2}{\line(0,-1){0.21}}
\multiput(0.87,49.3)(0.14,-0.2){2}{\line(0,-1){0.2}}
\multiput(1.15,48.89)(0.15,-0.2){2}{\line(0,-1){0.2}}
\multiput(1.44,48.49)(0.15,-0.2){2}{\line(0,-1){0.2}}
\multiput(1.74,48.09)(0.1,-0.13){3}{\line(0,-1){0.13}}
\multiput(2.05,47.7)(0.11,-0.13){3}{\line(0,-1){0.13}}
\multiput(2.36,47.32)(0.11,-0.12){3}{\line(0,-1){0.12}}
\multiput(2.69,46.95)(0.11,-0.12){3}{\line(0,-1){0.12}}
\multiput(3.03,46.58)(0.11,-0.12){3}{\line(0,-1){0.12}}
\multiput(3.37,46.22)(0.12,-0.12){3}{\line(0,-1){0.12}}
\multiput(3.72,45.87)(0.12,-0.11){3}{\line(1,0){0.12}}
\multiput(4.08,45.53)(0.12,-0.11){3}{\line(1,0){0.12}}
\multiput(4.45,45.19)(0.12,-0.11){3}{\line(1,0){0.12}}
\multiput(4.82,44.86)(0.13,-0.11){3}{\line(1,0){0.13}}
\multiput(5.2,44.55)(0.13,-0.1){3}{\line(1,0){0.13}}
\multiput(5.59,44.24)(0.2,-0.15){2}{\line(1,0){0.2}}
\multiput(5.99,43.94)(0.2,-0.15){2}{\line(1,0){0.2}}
\multiput(6.39,43.65)(0.2,-0.14){2}{\line(1,0){0.2}}
\multiput(6.8,43.37)(0.21,-0.14){2}{\line(1,0){0.21}}
\multiput(7.22,43.1)(0.21,-0.13){2}{\line(1,0){0.21}}
\multiput(7.64,42.84)(0.21,-0.13){2}{\line(1,0){0.21}}
\multiput(8.07,42.58)(0.22,-0.12){2}{\line(1,0){0.22}}
\multiput(8.5,42.34)(0.22,-0.12){2}{\line(1,0){0.22}}
\multiput(8.94,42.11)(0.22,-0.11){2}{\line(1,0){0.22}}
\multiput(9.39,41.89)(0.22,-0.11){2}{\line(1,0){0.22}}
\multiput(9.84,41.68)(0.23,-0.1){2}{\line(1,0){0.23}}
\multiput(10.29,41.48)(0.23,-0.1){2}{\line(1,0){0.23}}
\multiput(10.75,41.29)(0.46,-0.18){1}{\line(1,0){0.46}}
\multiput(11.21,41.11)(0.47,-0.17){1}{\line(1,0){0.47}}
\multiput(11.68,40.94)(0.47,-0.16){1}{\line(1,0){0.47}}
\multiput(12.15,40.78)(0.47,-0.15){1}{\line(1,0){0.47}}
\multiput(12.63,40.63)(0.48,-0.14){1}{\line(1,0){0.48}}
\multiput(13.1,40.5)(0.48,-0.12){1}{\line(1,0){0.48}}
\multiput(13.58,40.37)(0.48,-0.11){1}{\line(1,0){0.48}}
\multiput(14.07,40.26)(0.49,-0.1){1}{\line(1,0){0.49}}
\multiput(14.55,40.16)(0.49,-0.09){1}{\line(1,0){0.49}}
\multiput(15.04,40.07)(0.49,-0.08){1}{\line(1,0){0.49}}
\multiput(15.53,39.99)(0.49,-0.07){1}{\line(1,0){0.49}}
\multiput(16.03,39.92)(0.49,-0.06){1}{\line(1,0){0.49}}
\multiput(16.52,39.86)(0.49,-0.05){1}{\line(1,0){0.49}}
\multiput(17.01,39.81)(0.5,-0.03){1}{\line(1,0){0.5}}
\multiput(17.51,39.78)(0.5,-0.02){1}{\line(1,0){0.5}}
\multiput(18,39.76)(0.5,-0.01){1}{\line(1,0){0.5}}
\put(18.5,39.75){\line(1,0){0.5}}
\multiput(19,39.75)(0.5,0.01){1}{\line(1,0){0.5}}
\multiput(19.5,39.76)(0.5,0.02){1}{\line(1,0){0.5}}
\multiput(19.99,39.78)(0.5,0.03){1}{\line(1,0){0.5}}
\multiput(20.49,39.81)(0.49,0.05){1}{\line(1,0){0.49}}
\multiput(20.98,39.86)(0.49,0.06){1}{\line(1,0){0.49}}
\multiput(21.47,39.92)(0.49,0.07){1}{\line(1,0){0.49}}
\multiput(21.97,39.99)(0.49,0.08){1}{\line(1,0){0.49}}
\multiput(22.46,40.07)(0.49,0.09){1}{\line(1,0){0.49}}
\multiput(22.95,40.16)(0.49,0.1){1}{\line(1,0){0.49}}
\multiput(23.43,40.26)(0.48,0.11){1}{\line(1,0){0.48}}
\multiput(23.92,40.37)(0.48,0.12){1}{\line(1,0){0.48}}
\multiput(24.4,40.5)(0.48,0.14){1}{\line(1,0){0.48}}
\multiput(24.87,40.63)(0.47,0.15){1}{\line(1,0){0.47}}
\multiput(25.35,40.78)(0.47,0.16){1}{\line(1,0){0.47}}
\multiput(25.82,40.94)(0.47,0.17){1}{\line(1,0){0.47}}
\multiput(26.29,41.11)(0.46,0.18){1}{\line(1,0){0.46}}
\multiput(26.75,41.29)(0.23,0.1){2}{\line(1,0){0.23}}
\multiput(27.21,41.48)(0.23,0.1){2}{\line(1,0){0.23}}
\multiput(27.66,41.68)(0.22,0.11){2}{\line(1,0){0.22}}
\multiput(28.11,41.89)(0.22,0.11){2}{\line(1,0){0.22}}
\multiput(28.56,42.11)(0.22,0.12){2}{\line(1,0){0.22}}
\multiput(29,42.34)(0.22,0.12){2}{\line(1,0){0.22}}
\multiput(29.43,42.58)(0.21,0.13){2}{\line(1,0){0.21}}
\multiput(29.86,42.84)(0.21,0.13){2}{\line(1,0){0.21}}
\multiput(30.28,43.1)(0.21,0.14){2}{\line(1,0){0.21}}
\multiput(30.7,43.37)(0.2,0.14){2}{\line(1,0){0.2}}
\multiput(31.11,43.65)(0.2,0.15){2}{\line(1,0){0.2}}
\multiput(31.51,43.94)(0.2,0.15){2}{\line(1,0){0.2}}
\multiput(31.91,44.24)(0.13,0.1){3}{\line(1,0){0.13}}
\multiput(32.3,44.55)(0.13,0.11){3}{\line(1,0){0.13}}
\multiput(32.68,44.86)(0.12,0.11){3}{\line(1,0){0.12}}
\multiput(33.05,45.19)(0.12,0.11){3}{\line(1,0){0.12}}
\multiput(33.42,45.53)(0.12,0.11){3}{\line(1,0){0.12}}
\multiput(33.78,45.87)(0.12,0.12){3}{\line(1,0){0.12}}
\multiput(34.13,46.22)(0.11,0.12){3}{\line(0,1){0.12}}
\multiput(34.47,46.58)(0.11,0.12){3}{\line(0,1){0.12}}
\multiput(34.81,46.95)(0.11,0.12){3}{\line(0,1){0.12}}
\multiput(35.14,47.32)(0.11,0.13){3}{\line(0,1){0.13}}
\multiput(35.45,47.7)(0.1,0.13){3}{\line(0,1){0.13}}
\multiput(35.76,48.09)(0.15,0.2){2}{\line(0,1){0.2}}
\multiput(36.06,48.49)(0.15,0.2){2}{\line(0,1){0.2}}
\multiput(36.35,48.89)(0.14,0.2){2}{\line(0,1){0.2}}
\multiput(36.63,49.3)(0.14,0.21){2}{\line(0,1){0.21}}
\multiput(36.9,49.72)(0.13,0.21){2}{\line(0,1){0.21}}
\multiput(37.16,50.14)(0.13,0.21){2}{\line(0,1){0.21}}
\multiput(37.42,50.57)(0.12,0.22){2}{\line(0,1){0.22}}
\multiput(37.66,51)(0.12,0.22){2}{\line(0,1){0.22}}
\multiput(37.89,51.44)(0.11,0.22){2}{\line(0,1){0.22}}
\multiput(38.11,51.89)(0.11,0.22){2}{\line(0,1){0.22}}
\multiput(38.32,52.34)(0.1,0.23){2}{\line(0,1){0.23}}
\multiput(38.52,52.79)(0.1,0.23){2}{\line(0,1){0.23}}
\multiput(38.71,53.25)(0.18,0.46){1}{\line(0,1){0.46}}
\multiput(38.89,53.71)(0.17,0.47){1}{\line(0,1){0.47}}
\multiput(39.06,54.18)(0.16,0.47){1}{\line(0,1){0.47}}
\multiput(39.22,54.65)(0.15,0.47){1}{\line(0,1){0.47}}
\multiput(39.37,55.13)(0.14,0.48){1}{\line(0,1){0.48}}
\multiput(39.5,55.6)(0.12,0.48){1}{\line(0,1){0.48}}
\multiput(39.63,56.08)(0.11,0.48){1}{\line(0,1){0.48}}
\multiput(39.74,56.57)(0.1,0.49){1}{\line(0,1){0.49}}
\multiput(39.84,57.05)(0.09,0.49){1}{\line(0,1){0.49}}
\multiput(39.93,57.54)(0.08,0.49){1}{\line(0,1){0.49}}
\multiput(40.01,58.03)(0.07,0.49){1}{\line(0,1){0.49}}
\multiput(40.08,58.53)(0.06,0.49){1}{\line(0,1){0.49}}
\multiput(40.14,59.02)(0.05,0.49){1}{\line(0,1){0.49}}
\multiput(40.19,59.51)(0.03,0.5){1}{\line(0,1){0.5}}
\multiput(40.22,60.01)(0.02,0.5){1}{\line(0,1){0.5}}
\multiput(40.24,60.5)(0.01,0.5){1}{\line(0,1){0.5}}

\linethickness{0.2mm}
\qbezier(13.25,81.68)(13.11,81.6)(12.25,80.58)
\qbezier(12.25,80.58)(11.4,79.55)(11.49,78.61)
\qbezier(11.49,78.61)(11.86,77.87)(13.03,77.74)
\qbezier(13.03,77.74)(14.2,77.61)(14.56,76.86)
\qbezier(14.56,76.86)(14.63,76.03)(13.68,75.33)
\qbezier(13.68,75.33)(12.74,74.62)(12.81,73.79)
\qbezier(12.81,73.79)(13.17,73.04)(14.34,72.91)
\qbezier(14.34,72.91)(15.51,72.79)(15.88,72.04)
\qbezier(15.88,72.04)(15.94,71.21)(15,70.5)
\qbezier(15,70.5)(14.06,69.8)(14.12,68.97)
\qbezier(14.12,68.97)(14.49,68.22)(15.66,68.09)
\qbezier(15.66,68.09)(16.83,67.96)(17.19,67.21)
\qbezier(17.19,67.21)(17.26,66.38)(16.32,65.68)
\qbezier(16.32,65.68)(15.37,64.97)(15.44,64.14)
\qbezier(15.44,64.14)(15.8,63.39)(16.97,63.27)
\qbezier(16.97,63.27)(18.14,63.14)(18.51,62.39)
\linethickness{0.2mm}
\qbezier(-2.5,61.25)(-2.45,61.09)(-1.69,60)
\qbezier(-1.69,60)(-0.93,58.91)(0,58.75)
\qbezier(0,58.75)(0.82,58.91)(1.25,60)
\qbezier(1.25,60)(1.68,61.09)(2.5,61.25)
\qbezier(2.5,61.25)(3.32,61.09)(3.75,60)
\qbezier(3.75,60)(4.18,58.91)(5,58.75)
\qbezier(5,58.75)(5.82,58.91)(6.25,60)
\qbezier(6.25,60)(6.68,61.09)(7.5,61.25)
\qbezier(7.5,61.25)(8.32,61.09)(8.75,60)
\qbezier(8.75,60)(9.18,58.91)(10,58.75)
\qbezier(10,58.75)(10.82,58.91)(11.25,60)
\qbezier(11.25,60)(11.68,61.09)(12.5,61.25)
\qbezier(12.5,61.25)(13.32,61.09)(13.75,60)
\qbezier(13.75,60)(14.18,58.91)(15,58.75)
\qbezier(15,58.75)(15.82,58.91)(16.25,60)
\qbezier(16.25,60)(16.68,61.09)(17.5,61.25)
\linethickness{0.2mm}
\qbezier(38.89,69.27)(38.79,69.4)(37.68,70.14)
\qbezier(37.68,70.14)(36.57,70.88)(35.65,70.69)
\qbezier(35.65,70.69)(34.94,70.24)(34.94,69.07)
\qbezier(34.94,69.07)(34.94,67.89)(34.23,67.45)
\qbezier(34.23,67.45)(33.41,67.3)(32.61,68.16)
\qbezier(32.61,68.16)(31.81,69.02)(30.99,68.87)
\qbezier(30.99,68.87)(30.29,68.42)(30.28,67.25)
\qbezier(30.28,67.25)(30.28,66.07)(29.58,65.63)
\qbezier(29.58,65.63)(28.76,65.47)(27.96,66.34)
\qbezier(27.96,66.34)(27.16,67.2)(26.34,67.04)
\qbezier(26.34,67.04)(25.63,66.6)(25.63,65.42)
\qbezier(25.63,65.42)(25.62,64.25)(24.92,63.81)
\qbezier(24.92,63.81)(24.1,63.65)(23.3,64.51)
\qbezier(23.3,64.51)(22.5,65.38)(21.68,65.22)
\qbezier(21.68,65.22)(20.97,64.78)(20.97,63.6)
\qbezier(20.97,63.6)(20.97,62.43)(20.26,61.98)
\linethickness{0.2mm}
\qbezier(4.09,45.87)(4.24,45.79)(5.55,45.59)
\qbezier(5.55,45.59)(6.87,45.39)(7.63,45.95)
\qbezier(7.63,45.95)(8.08,46.65)(7.58,47.72)
\qbezier(7.58,47.72)(7.09,48.79)(7.54,49.49)
\qbezier(7.54,49.49)(8.22,49.97)(9.31,49.53)
\qbezier(9.31,49.53)(10.4,49.09)(11.07,49.57)
\qbezier(11.07,49.57)(11.53,50.27)(11.03,51.34)
\qbezier(11.03,51.34)(10.54,52.41)(10.99,53.11)
\qbezier(10.99,53.11)(11.67,53.59)(12.76,53.15)
\qbezier(12.76,53.15)(13.85,52.71)(14.52,53.19)
\qbezier(14.52,53.19)(14.98,53.89)(14.48,54.96)
\qbezier(14.48,54.96)(13.99,56.03)(14.44,56.73)
\qbezier(14.44,56.73)(15.12,57.21)(16.21,56.77)
\qbezier(16.21,56.77)(17.3,56.33)(17.97,56.81)
\qbezier(17.97,56.81)(18.43,57.51)(17.93,58.58)
\qbezier(17.93,58.58)(17.44,59.65)(17.89,60.35)
\linethickness{0.2mm}
\qbezier(34.29,46.86)(34.36,47.01)(34.5,48.34)
\qbezier(34.5,48.34)(34.65,49.66)(34.05,50.39)
\qbezier(34.05,50.39)(33.33,50.81)(32.28,50.27)
\qbezier(32.28,50.27)(31.24,49.72)(30.52,50.14)
\qbezier(30.52,50.14)(30,50.8)(30.4,51.91)
\qbezier(30.4,51.91)(30.79,53.02)(30.27,53.67)
\qbezier(30.27,53.67)(29.55,54.09)(28.51,53.55)
\qbezier(28.51,53.55)(27.47,53)(26.75,53.42)
\qbezier(26.75,53.42)(26.23,54.08)(26.62,55.19)
\qbezier(26.62,55.19)(27.01,56.3)(26.5,56.95)
\qbezier(26.5,56.95)(25.78,57.37)(24.74,56.83)
\qbezier(24.74,56.83)(23.69,56.28)(22.97,56.7)
\qbezier(22.97,56.7)(22.46,57.36)(22.85,58.47)
\qbezier(22.85,58.47)(23.24,59.58)(22.73,60.23)
\qbezier(22.73,60.23)(22,60.65)(20.96,60.11)
\qbezier(20.96,60.11)(19.92,59.56)(19.2,59.98)
\linethickness{0.2mm}
\put(18.75,61.25){\circle{3.54}}
\end{picture}
}

\renewcommand{\theequation}{\arabic{section}.\arabic{equation}}

\begin{document}
\renewcommand{\thefootnote}{\fnsymbol{footnote}}
\begin{flushright}
\vspace{1mm}
\end{flushright}

\vspace{1cm}

\begin{center}
{\bf \LARGE
Exact higher-spin symmetry in CFT: \\ %
\vspace{0.3cm} all correlators in unbroken Vasiliev theory
}

\vspace{2cm}

\textsc{V.E. Didenko\footnote{didenko@lpi.ru}
and E.D. Skvortsov\em${}^*$\footnote{skvortsov@lpi.ru}}

\vspace{2cm}

{\em${}^*$Lebedev Institute of Physics, Moscow, Russia }
\vspace*{.5cm}

{\em${}^\dag$  Albert Einstein Institute, Potsdam, Germany}

\vspace{1cm}
\end{center}

\vspace{0.5cm}
\begin{abstract}
All correlation functions of conserved currents of the CFT that is
dual to unbroken Vasiliev theory are found as invariants of
higher-spin symmetry in the bulk of AdS. The conformal and
higher-spin symmetry of the correlators as well as the
conservation of currents are manifest, which also provides a
direct link between the Maldacena-Zhiboedov result and higher-spin
symmetries. Our method is in the spirit of AdS/CFT, though we
never take any boundary limit or compute any bulk integrals.
Boundary-to-bulk propagators are shown to exhibit an algebraic
structure, living at the boundary of SpH(4), semidirect product of
Sp(4) and the Heisenberg group. N-point correlation function is
given by a product of N elements.
\end{abstract}

\newpage
\tableofcontents

\renewcommand{\thefootnote}{\arabic{footnote}}
\setcounter{footnote}{0}

\section{Introduction}\setcounter{equation}{0}
The conformal symmetry has been always important as the ultimate symmetry:
of UV and IR fixed points of QFT's,
as the symmetry of a dual to a theory of quantum gravity in AdS and for the
study of critical phenomena. The Virasoro symmetry combined with
the ideas of conformal bootstrap and OPE has already given a
harvest of exactly solvable models \cite{Belavin:1984vu}.
These are confined to two space-time dimensions though.

To find exactly solvable models in higher dimensions one calls for
an infinite symmetry enveloping the conformal algebra, for the
conformal symmetry itself seems to be too weak in $d>2$.
Higher-spin symmetry (HS) could provide a proper replacement for
the Virasoro one. However, in $d>2$ HS symmetry, if unbroken,
turns out to be too restrictive for a CFT to have interactions as Maldacena and Zhiboedov
have shown recently \cite{Maldacena:2011jn}. The 'minimal models' with exact HS
symmetry are free theories. It is
tempting to say that nontrivial 'minimal models' in $d>2$ result from
breaking HS symmetries. Indeed, the results of
\cite{Maldacena:2012sf} indicate that broken HS symmetry is still
quite restrictive. 

In this paper we apply HS symmetry to calculate all correlation
functions of a CFT that has an exact HS symmetry. The correlators are found
for the operators that are dual to HS gauge fields in $AdS$. We
have chosen the simplest HS algebra that underlies the Vasiliev HS
theory in $AdS_4$. The methods that we shall use in this paper for calculating correlation
functions have been spelled out in \cite{Colombo:2012jx} and follow the general prescription of AdS/CFT, \cite{Maldacena:1997re, Gubser:1998bc, Witten:1998qj}.
However, at no stage we need to take the boundary limit
$z\rightarrow0$ or deal with integrals over the bulk of $AdS$. The results are manifestly independent of any coordinate choice in the bulk. The point of view on holography at
any $z$ is natural within the Vasiliev approach to HS and has been laid
already in \cite{Vasiliev:2012vf}.

The HS symmetry of the correlators as well as the conservation of the currents are manifest.
Therefore, the results of the paper can be also viewed as a complement to the Maldacena-Zhiboedov
theorem \cite{Maldacena:2011jn}. To be more precise, \cite{Maldacena:2011jn} classifies all HS algebras together with their unitary irreducible representations, under certain assumptions on the spectrum of
conformal primaries. The latter determines the form of the correlators. We assume the existence of some HS algebra and use it to directly determine all the correlation functions.

In addition, the Maldacena-Zhiboedov result \cite{Maldacena:2011jn} is quite technical in its proof and it would be instructive to have a direct link between HS symmetries and correlation functions. This is
necessary in order to study the breaking of HS symmetries. Providing such a link
is also a result of the present paper. We show that HS symmetry transformations can be viewed either as bulk symmetries or as boundary symmetries, the latter containing conformal algebra as a subalgebra.

The results can be generalized to any free CFT and it would be interesting to find correlation functions
of free $\mathcal{N}=4$ SYM, reveal the implications of a slightly broken HS symmetry and see if it continues to persist up to the strong coupling.

Higher-spin (HS) theory at present is given by Vasiliev \cite{Vasiliev:1990en, Vasiliev:1990vu, Vasiliev:1992av, Vasiliev:1995dn, Vasiliev:1999ba, Vasiliev:2003ev} in the form of classical equations of motion in the bulk of $AdS$ space
that are invariant under diffeomorphisms and HS gauge transformations and whose linearization describes
propagation of free fields of all spins $s=0,1,2,3,...$, there is also a truncation to even spins $s=0,2,4,...$. The HS fields are gauge fields, being totally-symmetric as tensors,
\be \delta\phi_{\mu_1...\mu_s}=\pl_{\mu_1} \xi_{\mu_2...\mu_3} +\mbox{permutations}\,,\ee
hence these should be $AdS/CFT$ dual to conserved currents
\be j_{i_1...i_s}\,, \qquad\qquad\pl^k j_{k i_2...i_s}=0\,,\qquad\qquad \Delta=s+1\,,\qquad s>0\ee A natural candidate for the CFT dual was conjectured to be a free vector model, \cite{Sezgin:2002rt, Klebanov:2002ja}, which has the same spectrum of singlet currents
\be j_{k(s)}=\phi(x)(\overleftarrow{\pl_k}-\overrightarrow{\pl_k})^s \phi(x)\,,\ee
where $\phi(x)$ is a vector multiplet of free scalars. Analogous set of currents can be constructed out  of free fermion, with the only difference in the conformal weight of $j_0=\phi^2$, which is $1$ or $2$, respectively. At free level the AdS/CFT duality is essentially the Flato-Fronsdal theorem \cite{Flato:1978qz}.

The Flato-Fronsdal theorem states, in its conformal setting, that the tensor square of a free conformal scalar, i.e. the bilocal operator $\phi(x)\phi(y)$,
decomposes as a representation of the conformal algebra into a direct sum of conserved currents $j_s$ plus $j_0=\phi^2(x)$. We can rephrase this as the OPE in terms of the conformal algebra primaries
\be\phi\times \phi=\mathbf{1}+\sum_s j_s\,\ee
where $\mathbf{1}$ is the identity operator. A more compact way to state this result is to use the OPE of the fields that are primaries of the HS algebra
\be\phi\times\phi=\mathbf{1}+\mathbb{J}\ee
where the free conformal scalar $\phi$, the identity operator $\mathbf{1}$ and the direct sum $\mathbb{J}=\sum_s j_s$ of the currents are unitary representations of the HS algebra. The latter is reducible under the conformal algebra. Analogous statements are valid for the free fermion theory. In this way  HS symmetry allows one to combine an infinite number of the conformal algebra primaries into a finite number of the HS algebra primaries, the job done by the Virasoro algebra in $2d$.

Within the $AdS/CFT$ paradigm Vasiliev higher-spin theory has to
be supplemented with certain boundary conditions at the conformal
infinity of $AdS$. Generic boundary conditions break higher-spin
symmetries. For a special choice of boundary behavior the
higher-spin symmetry seems to remain unbroken to all orders in
perturbation theory \cite{Vasiliev:2012vf}, implying the boundary
theory is free by the Maldacena-Zhiboedov theorem
\cite{Maldacena:2011jn}, i.e. all correlation functions are given
either by free boson or by free fermion.

Despite the powerful result of \cite{Maldacena:2011jn} on the CFT
side, the precise mechanism in the bulk is not fully understood
yet due to the complexity of the bulk theory. The crucial tests of
the conjecture have been performed in \cite{Giombi:2009wh,
Giombi:2010vg} where three-point functions have been computed
directly from Vasiliev theory by solving bulk equations to the
second order with two current sources on the boundary and then
taking the boundary limit of the solution. The computations of
\cite{Giombi:2009wh, Giombi:2010vg} were quite involved with a
simple result in the end, moreover some ambiguities to be
resolved have been observed at the intermediate steps.

On the other hand, the formalism of Vasiliev theory suggests
introducing observables, \cite{Sezgin:2005pv}, the
quantities built of HS master fields that are by definition
invariant under all HS symmetries provided that they exist in
sense of producing a finite result. The simplest such observables
are (long)trace operators schematically given by
\be
O_n=\mbox{Tr}(\Phi\star...\star \Phi),\qquad\qquad  \delta\Phi=[\Phi,\xi]_\star\,,
\ee
where it is important that $\Phi$, which is related to HS master
field strength, transforms in the adjoint representation of the HS
algebra, which is realized as certain $\star$-product, and the
existence of trace operation is also assumed. In this case $O_n$
is invariant under all HS transformations. The existence of such invariants was first pointed out in \cite{Vasiliev:1986qx}. If $\Phi$ is related to
boundary-to-bulk propagators, i.e. it depends on the boundary point $\rmx$ and bulk point $X$, $\Phi=\Phi(X, \rmx)$, with the fall-off conditions
preserved by HS symmetry, $O_n$ should correspond to the correlation
functions of conserved currents
\be
\langle j(\rmx_1)...j(\rmx_n)\rangle=\sum_{S_n}\mbox{Tr}(\Phi(X,\rmx_1)\star...\star \Phi(X,\rmx_n))\,. \label{allcorrfunc}
\ee
Crucial is that the dependence on the bulk point $X$ drops out of
\eqref{allcorrfunc} as the change of $X$ is a particular large
HS transformation. This way of extracting correlation functions
was proposed by Colombo and Sundell in \cite{Colombo:2012jx} and
confirmed at the level of $2$- and $3$-point functions. Within the
general framework of \cite{Sezgin:2011hq,Colombo:2010fu,
Sezgin:2012ag, Colombo:2012jx} applied in \cite{Colombo:2012jx}
there appear to be several types of divergences that need to be
regularized, leading to quite a complicated technique. However,
unlike well-defined star-product trace that is used in our paper,
the very definition of the trace in \cite{Colombo:2012jx} requires
regularization.

The goal of the present paper is to show that (\ref{allcorrfunc})
is a perfectly finite quantity that can be easily computed using
the algebraic structure of $\Phi$ that we found. It turns out that
the boundary-to-bulk propagators posses certain projective
properties within the $\star$-product. In particular the closed form
for all $n$-point correlation functions of conserved currents can
be easily obtained. This calculation is carried out at the free level
of HS equations in four dimensions. The nonlinear contribution for
appropriate boundary conditions that do not break HS symmetry is
believed to leave the result for correlation functions unaffected
in accordance with \cite{Klebanov:2002ja, Maldacena:2011jn, Vasiliev:2012vf, Colombo:2012jx}.

The other result that lies aside from primary goal of this paper
is the group-theoretical approach to certain $\star$-product elements
that arise in the calculation of correlation functions. The
boundary-to-bulk propagators turn out to belong to a subalgebra
in HS algebra of the elements ($Y\equiv Y^A$ are auxiliary
$sp(4)\sim so(3,2)$ spinors with the Weyl $\star$-product
realization of the HS algebra) that are Gaussians in $Y$
\be
\Phi(f^{AB},\xi^A, a)=\exp i \left\{\frac12 Yf\,Y+\xi\, Y +a
\right\}\,,\label{genbtob}
\ee
where $f$ and $\xi$ depend on $X$ and $\rmx$.
The product of $n$ such elements naturally springs up in our
calculations. We show that subalgebra \eqref{genbtob} can be
conveniently parameterized by the $SpH(4)$ group. In other words,
each of elements \eqref{genbtob} and their $\star$-product belong to
$SpH(4)$ -- the semidirect product  of $Sp(4)$ and the Heisenberg
group $(\xi,a)\star (\eta,b)=(\xi+\eta, a+b+\xi\eta)$, where
central elements $a$, $b$ are just numbers and $\xi^A$ and $\eta^A$
are contracted with the $Sp(4)$ invariant metric. The trace operation projects
onto the central part. To
be more precise, the quadratic form in $\Phi(f,\xi,a)$
parameterizes the Lie algebra of $SpH(4)$, while the group
elements are obtained via the Cayley transform. HS propagators
turn out to be even more peculiar as they are not generic elements
of type (\ref{genbtob}), lying in the subspace where Cayley
transform is not invertible but the $\star$-product is still
well-defined. Finally, $n$-point correlation function is a
centrally projected product of $n$ such propagators (\ref{genbtob})

The minimal HS prerequisites we need to discuss the constructive
approach to correlation functions invariant under unbroken HS
symmetry are given in Section \ref{sec:prereq}. In Section
\ref{sec:btobproperties} we discuss boundary-to-bulk propagators.
The observables are introduced in Section \ref{sec:Observables}. The explicit relation between
propagators and $SpH(4)$ is presented in Section
\ref{sec:btotbGroup}.

The conformal and HS symmetry of the $n$-point function is
manifest. Nevertheless, we consider it important to explicitly
expand the result in terms of conformal invariants introduced in
\cite{Costa:2011mg, Giombi:2011rz}. Such rewriting manifests
the conformal symmetry of the result but not its HS symmetry and the
conservation of the currents is also hard to see. This is done in
Section \ref{sec:Npoints} where the examples of $2$, $3$ and $4$-point
functions are considered in detail.  The main result is the closed form for the
$n$-point function (\ref{npointAnswer}). Conclusion is given in
Section \ref{sec:Conclusion}.

\section{HS prerequisites}\setcounter{equation}{0}\label{sec:prereq}
The higher-spin equations are formulated in terms of certain
master fields. There are two of them, a master HS connection
$\omega$, which contains gauge fields themselves with certain
their derivatives, and a master field strength $B$, which contains
gauge invariant field strengths together with all their
on-mass-shell nontrivial derivatives. The lesson of HS is that when fields are supplemented with certain derivatives
thereof to form a master field an extended symmetry acts in a
simple way on the master field.

\paragraph{Higher-spin algebra.}
Crucial for the whole $AdS_4$ HS story is the isomorphism $so(3,2)\sim sp(4,\mathbb{R})$ so we will use $sp(4)$ instead. The $sp(4)$ generators $T_{AB}=T_{BA}$, $A,B,...=1...4$ obey
\be[T_{AB},T_{CD}]=T_{AD}\epsilon_{BC}+\mbox{3 terms}\,,\ee
where $\epsilon_{AB}=-\epsilon_{BA}$,
$\epsilon_{AB}\epsilon^{AC}=\delta\fdu{B}{C}$ is a symplectic
form. Splitting $A=\{\ga,\gad\}$, etc. along with the choice
$\epsilon_{AB}=\mbox{diag}(\epsilon_{\ga\gb},\epsilon_{\gad\gbd})$,
$\epsilon_{\ga\gb}=-\epsilon_{\gb\ga}$, $\epsilon_{12}=1$ {\em
idem.} for $\epsilon_{\gad\gbd}$, the generators of the bulk
Lorentz $so(3,1)\sim sp_2\oplus sp_2$ algebra are given by
$T_{\ga\gb}$, $\bar{T}_{\gad\gbd}$, and  $T_{\ga\gad}$ are the
$AdS_4$ translations.

The higher-spin algebra \cite{Fradkin:1986ka, Vasiliev:1986qx} is
a universal enveloping algebra of $sp_4$ quotiented by the
two-sided ideal that is the singleton annulator,
\cite{Konshtein:1988yg, Konstein:1989ij}. This is the point where
the singleton, i.e. free $3d$ conformal scalar/fermion enters the story. In
the case of $so(3,2)$ the ideal is resolved by an oscillator realization
of $sp_4$. Namely, elements of the higher-spin algebra are
functions of formally commuting variables $Y_A$ with the Weyl
product
\be\label{star}
f(Y)\star g(Y)=\int\, dU\,dV\, f(Y+U)g(Y+V)e^{iV^{A}
U_{A}}=f(Y)\exp
\left\{i\overleftarrow{\pl_A}\epsilon^{AB}\overrightarrow{\pl_B}
\right\}g(Y)\,,\quad
\ee
($\epsilon_{AB}$ is used to raise and lower indices $Y^A=\epsilon^{AB}Y_B$, $Y_A=Y^B\epsilon_{BA}$), which effectively implies
\be
[Y_A,\,Y_B]_\star=2i\epsilon_{AB}\,, \qquad\quad Y_A\star f(Y)=Y_A
f+i\frac{\p f}{\p Y^A}\,,\qquad f(Y)\star Y_{A}=Y_A f-i\frac{\p
f}{\p Y^A}\,.
\ee
The $sp_4$ generators, which form a Lie subalgebra under the
$\star$-commutator, read
\be\label{spfour}
T_{AB}=-\frac{i}{4}\{Y_A,Y_B\}_\star\,, \qquad\qquad [T_{AB},Y_C]=Y_A\epsilon_{BC}+Y_B\epsilon_{AC}\,.
\ee

\paragraph{Linearized equations.}
The full nonlinear Vasiliev equations is a subject of several
reviews,
\cite{Vasiliev:1995dn,Vasiliev:1999ba,Bekaert:2005vh,Sezgin:2012ag},
see also original works \cite{Vasiliev:1990en, Vasiliev:1990vu,
Vasiliev:1992av, Vasiliev:1995dn, Vasiliev:1999ba,
Vasiliev:2003ev}. Here we need the linearized equations only, a
part of them actually that encode HS field strengths. These are
contained in the master field $B(Y|X)$, which is a space-time zero-form. Projection onto the integer spin fields imposes
kinematic constraint $B(Y)=B(-Y)$. Various components of $B(Y)$ at
the free level are identified as follows,
$$B(Y|X)=\sum B_{A(k)}(X)\,Y^{A(k)}=\sum_{k,m} B_{\ga(k),\gad(m)}(X)\,y^{\ga(k)}\,\bar{y}^{\gad(m)}$$
\vspace{0.2cm}
\noindent\begin{tabular}{|c|w{9.2cm}|}
  \hline component & \centering meaning\tabularnewline\hline
 \rule{0pt}{12pt} $B$ & the scalar field \tabularnewline
  \rule{0pt}{12pt} $B_{\ga(k),\gad(k)}$,\,\,\, $k>0$ & derivatives of the scalar field $B$, $B_{\ga\gad}=D_{\ga\gad}B$, etc. \tabularnewline
  \rule{0pt}{12pt}$B_{\ga\gb}$\quad $B_{\gad\gbd}$ & (anti)selfdual parts of the Maxwell spin-one tensor $F_{\mu\nu}$\tabularnewline
 \rule{0pt}{12pt} $B_{\ga(4)}$\quad $B_{\gad(4)}$ & (anti)selfdual parts of the spin-two Weyl tensor \tabularnewline
  \rule{0pt}{20pt} $B_{\ga(2s)}$\quad $B_{\gad(2s)}$ & (anti)selfdual parts of the field strength for a spin-$s$ field, which are also called spin-$s$ Weyl tensors \tabularnewline
  \rule{0pt}{12pt} $B_{\ga(2s+k),\gad(k)}$ $B_{\ga(k),\gad(2s+k)}$,\, $k>0$ & derivatives of the spin-$s$ field strength\tabularnewline
  \hline
\end{tabular}
\vspace{0.2cm}

The equations for the master fields are first order differential equations that express exterior derivative of all
the fields in terms of exterior products of the fields themselves and are called unfolded equations, \cite{Vasiliev:1988xc, Vasiliev:1988sa}. The full Vasiliev equations have the unfolded form too. The linearized equations for the master field $B$ read
\begin{align}
d\Omega+\Omega\star\Omega&=0\label{UnfldA}\,,\\
dB+\Omega\star B-B\star \tilde{\Omega}&=0\,,\label{UnfldB}
\end{align}
where $d$ is de Rham differential, $\Omega=\frac12
\Omega^{AB}_\mu dx^\mu T_{AB}$ is a flat $sp_4$ connection that
contains vierbein $h^{\ga\gad}=\Omega^{\ga\gad}$ as well as
(anti)selfdual parts of the spin-connection, $\Omega^{\ga\gb}$,
$\bar\Omega^{\gad\gbd}$. $\tilde{\Omega}$ represents the action of
an $sp_4$ automorphism that flips the sign of $AdS_4$ translations
$\tilde{T}_{\ga\gad}=-T_{\ga\gad}$. $\Omega$ is a vacuum value of the
master field $\omega$ that contains all HS gauge fields, including
graviton as spin-two. It is only the graviton part of $\Omega$ that is nonzero.
Its purpose is to define $AdS_4$ background.

Eq. (\ref{UnfldB}) does several things: it expresses all
components $B_{\ga(2s+k),\gad(k)}$, $B_{\ga(k),\gad(2s+k)}$, with
$k>0$ as rank-$k$ derivatives of Weyl tensors $B_{\ga(2s)}$,
$B_{\gad(2s)}$; imposes Klein-Gordon equation on the scalar
$B(X)=B(0|X)$, $(\square -2)B(X)=0$; imposes Bianchi identities
and equations of motion, $D^{\gb\gbd}B_{\ga(2s-1)\gb}=0$,
$D^{\gb\gbd}B_{\gad(2s-1)\gbd}=0$, which are equations restricting
the form of Weyl tensors and are automatically satisfied once the Weyl
tensor is expressed as order-$s$ derivative of the gauge
potential.

\paragraph{Reality conditions.} $\star$-product \eqref{star} defined
on Majorana spinors $Y_{A}$ admits an involution
$y_{\al}^{\dagger}=\bar{y}_{\dal}$,
$\bar{y}_{\dal}^{\dagger}=y_{\al}$ such that
\be
(\gl f+\mu g)^{\dagger}=\bar{\gl}f^{\dagger}+\bar{\mu}
g^{\dagger}\,,\qquad (f\star g)^{\dagger}=g^{\dagger}\star
f^{\dagger}\,.
\ee
The appropriate reality conditions for background one-form $\gO$ and
zero-form $B$ read
\be
\gO^{\dagger}=-\gO\,,\qquad B^{\dagger}=\tilde{B}\,.
\ee

\paragraph{Symmetries and general solution.} Eqs. (\ref{UnfldA})-(\ref{UnfldB}) are invariant under the global HS symmetries
\begin{align}
\delta\Omega&=d\xi+[\Omega,\xi]_\star=0\,,\label{UnfldAA}\\
\delta B&=-\xi\star B+B\star \tilde{\xi}\,,\label{UnfldAB}
\end{align}
where the first equation imposes the invariance of the vacuum
$\delta \Omega=0$. Eq. (\ref{UnfldAB}) illustrates why it is
useful to pack fields together with their derivatives into a
master field as the extended symmetry, which is the HS symmetry,
gets explicitly algebraic.

Since $\Omega$ is flat, (\ref{UnfldAA}), one can represent it in the pure gauge form
\be\Omega=g^{-1}\star dg\,,\label{flatConnection}\ee
where $g=g(Y|X)$ is actually Gaussian $g\sim\exp i (\frac12
Y^A f\fdu{A}{B}Y_B)$ since $\Omega$ occupies only the $sp_4$ part
of the HS algebra and $T_{AB}$, (\ref{spfour}), are quadratic in $Y$. Then one can
solve (\ref{UnfldB}) and (\ref{UnfldAA})
\begin{align}
\xi(Y|X)&=g^{-1}\star\xi(Y|X_0)\star g\,,\label{largeTAdRotaionXi}\\
B(Y|X)&=g^{-1}\star B(Y|X_0)\star \tilde{g}\,,\label{largeTAdRotaion}
\end{align}
where $X_0$ is a point where $g(Y|X_0)=1$. On the other hand, one
can take arbitrary functions $\xi(Y)$ and $B(Y)$ that do not
depend on $X$ and obtain the solutions to (\ref{UnfldB}) and
(\ref{UnfldAA}) via (\ref{largeTAdRotaionXi}) and
(\ref{largeTAdRotaion}). Then the initial data $\xi(Y)$ and $B(Y)$
turn out to be equal to the solutions at $X_0$ where $g(X_0)=1$.
The Cauchy problem is naturally replaced by a Taylor-like problem
as $B(Y)$ parameterizes all nontrivial derivatives at a point.

That the solutions to (\ref{UnfldAA}) are parameterized
by $\xi(Y)$, which is an element of the HS algebra, implies that
the HS algebra is the algebra of global symmetries of
(\ref{UnfldA})-(\ref{UnfldB}).

\section{Boundary-to-bulk propagators} \setcounter{equation}{0}\label{sec:btobproperties}

Essential ingredient of the following construction is the
boundary-to-bulk propagator, $B(X|Y|\mathrm{x},\eta)$, which by
definition has two legs, one behaving as a bulk HS master field
$B$, i.e. a generating function of HS field strengths, and the
second leg behaving as a generating function for all conserved
currents on the boundary. It depends on the bulk coordinate $X$,
which in Poincare coordinates splits as $X=(z,x)$,
$x=x^{\ga\gb}$, where $x^{\al\gb}=x^{\gb\al}$ is the bi-spinor counterpart
to $x^{i}$; on auxiliary bulk variables $Y$
that allows us to pack all the field strengths together with
their derivatives into a single master field $B$; on boundary
coordinate $\mathrm{x}\equiv\mathrm{x}^{\ga\gb}$ and on boundary
polarization spinor $\eta\equiv\eta^\ga$ that pretty much as $Y$
is used to pack up all conserved currents
\be j(\mathrm{x},\eta)=\sum_s j_{\ga(2s)}\,\eta^{\ga}...\eta^{\ga}\,.\ee
Note that $j(\rmx,0)$ is not a current as it has no indices at all. It is dual to the scalar field of the HS multiplet i.e. $B(0|X)$.
Propagator does not depend on any other quantities but described above, it satisfies (\ref{UnfldB}) in the bulk and behaves as a conserved current on the boundary \cite{Giombi:2012he}
\be\frac{\pl^2}{\pl\eta^\ga \pl\eta^{\gb}}\frac{\pl}{\pl\mathrm{x}_{\ga\gb}}B(X,Y|\mathrm{x},\eta)=0\,.\label{btobConservation}\ee

In \cite{Giombi:2009wh}\footnote{As the conventions of
\cite{Giombi:2009wh} are different from ours and perhaps not fully
clear, we redo this HS exercise in Appendix \ref{sec:btob}.} the
propagator was found in Poincare coordinates,
$B(X,Y|\mathrm{x},\eta)=B(x,z;y,\bar{y}|\rmx_i,\eta)$.  The HS
observables and correlation functions do not depend on any
particular choice of bulk coordinates as will become clear soon.
The expressions below are given for clarity and to illustrate
certain general properties of propagators.

\paragraph{Background in Poincare coordinates. }
To proceed, it is convenient to use $AdS_4$ Poincare coordinates
\be
ds^2=\ff{1}{z^2}(dz^2+dx_i dx^i)\,,
\ee
where the 3d boundary coordinates $x^{i}$ have indices contracted
with the flat Minkowski metric $\eta_{ij}$. The components of the
background connection $\Omega$ read
\begin{align}\label{flatP}
\Omega^{\ga\ga}&=\frac{i}{2z}dx^{\ga\ga}\,, &
\Omega^{\gad\gad}&=-\frac{i}{2z}dx^{\gad\gad}\,,  &
\Omega^{\ga\gad}&=\frac{1}{2z}(-dx^{\ga\gad}+i\epsilon^{\ga\gad}dz)\,,
&
\end{align}
where we introduced the mixed epsilon-symbol
$\gep_{\al\dgb}=-\gep_{\dgb\al}$ to single out $z$-direction.
$x^{\al\dal}$ should not be confused with the $4d$ coordinates as
we adopt the following convention
\be
x_{\al\dal}=x_{\al}{}^{\gb}\gep_{\gb\dal}\,,\qquad
\gep_{\al\dgb}\gep^{\gb\dgb}=\delta_{\al}{}^{\gb}\,.
\ee
The gauge function $g(Y|x,z)$ that reproduces connection
\eqref{flatP} has the factorized form
\be\label{gPoincare}
g=g_p\star g_z\,,\qquad g_{p}=e^{\ff{i}{2}P_{\al\gb}
x^{\al\gb}}\,,\qquad
g_{z}=\ff{4\sqrt{z}}{(1+\sqrt{z})^2}e^{\ff{1-\sqrt{z}}{1+\sqrt{z}}\bar{y}_{\al}y^{\al}}\,,
\ee
where $P_{\al\gb}=i y^{-}_{\al}y_{\gb}^{-}$, $y^{-}_{\al}=\ff12
(\bar{y}_{\al}-iy_{\al})$ corresponds to the boundary Poincare
translations \cite{Vasiliev:2012vf}. The $AdS_4$ connection
$\Omega$ then reads
\be\Omega=g_z^{-1}\star\left(g_p^{-1}\star d g_p\right)\star g_z+g_z^{-1}\star dg_z\,,\ee
where the piece $g_p^{-1}\star d g_p$ is the flat Poincare (boundary) connection in Cartesian coordinates, i.e. $\frac{i}2P_{\ga\gb} dx^{\ga\gb}$. Note, that
$g(Y|x_{\al\gb}=0, z=1)=1$.

\paragraph{Bulk-to-boundary propagators.} The HS boundary-to-bulk propagator is found to be\footnote{\label{ft:BiField} The boundary-to-bulk propagator is actually a bi-field satisfying the same equations (\ref{UnfldA})-(\ref{UnfldB})
on the boundary and in the bulk. Indeed, (\ref{UnfldA})-(\ref{UnfldB}) are background independent and instead of
taking $AdS_4$ with a non-degenerate $sp_4$ connection $\Omega$ one may study them over the $3$-dimensional boundary, where a natural connection is given by Cartesian coordinates $g_p^{-1}\star dg_p$. Then one can show,
see also \cite{Vasiliev:2012vf}, that (\ref{UnfldA})-(\ref{UnfldB}) describe conserved currents. We will not
use this fact in the paper, so $\rmx$ and $\eta$ are just external parameters rather than variables analogous to $X$ and $Y$.}
\begin{align}\label{PropagatorPoincare}
B=K \exp i\{ -y F\bar{y}+\xi y +\theta \}+K \exp i\{ -y F\bar{y}+\bar{\xi} \bar{y} -\theta \}+\left(\substack{\xi\leftrightarrow-\xi\\ \bar{\xi}\leftrightarrow-\bar{\xi}}\right)\,,
\end{align}
where $y F\bar{y}\equiv y^\ga F\fdu{\ga}{\gad}\bar{y}_{\gad}$, $\xi
y=\xi^{\al}y_{\al}$, $\theta$ is an arbitrary constant and $K$ is
the $\Delta=1$ scalar boundary-to-bulk propagator,
\begin{align}
K=\frac{z}{(x-\rmx)^2+z^2}\,.
\end{align}
$F\equiv F^{\ga\gad}$ is $\pl^{\ga\gad} \ln K$ up to some factor
\begin{align}\label{F}
F^{\ga\gad}&=-\left(\frac{2z}{(x-\rmx)^2+z^2}(x-\rmx)^{\ga\gad}+
\frac{(x-\rmx)^2-z^2}{(x-\rmx)^2+z^2}\,i\epsilon^{\ga\gad}\right)\,.
\end{align}
A particular spin-$s$ Weyl tensor propagator is encoded in
$B_{\al(2s)}$ and $B_{\dal(2s)}$ components of
\eqref{PropagatorPoincare}. Bulk polarization spinors $\xi$ and
$\bar{\xi}$ represent the boundary polarization spinor $\eta^\ga$
parallel transported to the bulk point $(x,z)$ with the parallel
transport bispinor $\Pi^{\ga\gb}$
\begin{align}\label{Xidef}
\xi^\ga&= \Pi^{\ga\gb}\eta_\gb\,,&& \Pi^{\ga\gb}=K \left(\frac{1}{\sqrt{z}}\,(x-\rmx)^{\ga\gb}-\sqrt{z}\,i\epsilon^{\ga\gb}\right)\,, && \bar{\xi}^{\gad}=(\xi^\ga)^\dag\,.
\end{align}
The symmetrization in \eqref{PropagatorPoincare} over $\xi$, $-\xi$, {\em{idem.}} for
$\bar{\xi}$,  projects onto the
bosonic part $B(Y)=B(-Y)$. Parameter $\theta$ is introduced for
convenience, it is a free parameter in the full Vasiliev
equations. At $\theta=0$ and $\theta=\pi/2$ the theory is
parity-invariant, \cite{Vasiliev:1992av, Sezgin:2003pt}. And the
Vasiliev theory at $\theta=0$ and $\theta=\pi/2$ with boundary
conditions preserving HS symmetry is conjectured to be dual to
the free vector model, bosonic and fermionic,
respectively. For other values of $\theta$ any boundary conditions
break HS symmetry, \cite{Vasiliev:2012vf}, and there is also a
proposal for its dual \cite{Giombi:2011kc,Chang:2012kt}. The fields of the vector model
are vectors of some group, e.g. $O(N)$ or $U(N)$. In the case of
$O(N)$ all odd-spin singlet currents vanish unless there are
other flavor groups. The pure $O(N)$ model should be dual to the
minimal bosonic Vasiliev theory,  which contains fields of even
spins $s=0,2,4,...$ . $U(N)$ vector model or $O(N)$ model with
additional flavors possesses currents of all integer spins and should
be dual to the bosonic Vasiliev theory with spectrum
$s=0,1,2,3,...$. For
brevity we will always refer to vector model without specifying the group.

\paragraph{Invariant properties of propagators.} Given some particular coordinates it might not be clear
which properties of the propagator are invariant of a particular
coordinate choice. Here we wish to collect those that are
coordinate independent. First, $F^{\ga\gad}$ is a projective $Sp(2)$
parameterization of the conformal boundary
\begin{align}
F^{\ga\gad}(\lambda x,\lambda z,\lambda \rmx)&=F(x,z, \rmx)\,,  & \Pi(\lambda
x,\lambda z,\lambda \rmx)&=\lambda^{-\frac12}\Pi(x,z,\rmx)\,.
\end{align}

A distinguishing property of the HS propagators found in
\cite{Didenko:2012vh} for arbitrary $d$ which has been recently confirmed for
$d=3$ in \cite{Kraus:2012uf} is that in all cases the solution is based on
certain projectors within $\star$-product algebra. This fact as it
seems is closely related to the appearance of $\gd$-like sources
in the space-time equations. Their reincarnation results in
projectors on the twistor-space side. These can be thought of as
analogs of distributions in the $\star$-product algebra. As it happens to distributions not all of them can be multiplied, see \cite{Soloviev:2012} for more detail. Fortunately we will not meet these subtleties.

To be more specific, let us introduce the $\star$-product
projector
\be\label{projectorS}
P_0=\exp i{(- yF\bar{y}+\xi y)}\,,\qquad P_0\star P_0=P_0\,,
\ee
which requires
\be
F^{\ga\gad}F\fud{\gb}{\gad}=\epsilon^{\ga\gb}\,,\qquad
F^{\ga\gad}F\fdu{\ga}{\gbd}=\epsilon^{\gad\gbd}\,.\label{FisSpTwo}
\ee
The latter is an invariant property for any choice of bulk coordinates, in particular
it holds for \eqref{F}.
Note, $P_0$ is annihilated by two self-commuting oscillators
\be y^{\pm}_\al= F_{\al}{}^{\dgb}\bar{y}_{\dgb}\pm y_{\al}+\xi_{\al}\ee as follows
\be
y^{+}\star P_0=P_0\star y^-=0\,,\qquad P_0\sim
\gd(y^{+})\star\gd(y^-)\,,
\ee
which makes it clear e.g., that $P_0\star\tilde{P}_0$ does not
exist. In other terms the $\star$-product of two propagators from
boundary point $\rmx$ and its inverse-reflected point $-\rmx/\rmx^2$ diverges,
which as become clear soon is related to the singularity of
correlation functions at coincident points.

Anti-holomorphic projector
\be\label{projectorS}
\bar{P}_0=\exp i{(- y F\bar{y}+\bar{\xi}\bar{y})}\,,\qquad
\bar{P}_0\star\bar{P}_0=\bar{P}_0
\ee
is annihilated by the same $y^{\pm}_\al$ provided that $\xi=-F\bar{\xi}$, which is also
an invariant property.

\paragraph{Interplay between bulk and boundary global symmetries.} We would like to show that global $Sp(4)$ symmetries
can be treated either as symmetries of the bulk theory or as
symmetries of the boundary CFT. The difference between the bulk
and the boundary is that $Y^A$ variables are explicitly affected by
global $Sp(4)$ transformations inducing certain action in the
bulk, while boundary coordinate $\rmx$ and polarization spinor
$\eta$ remain untouched.

Let us be given some gauge function $g=g(X|Y)$ that leads to $\Omega$ via (\ref{flatConnection}) and hence performs certain $Sp(4)$ rotation of $Y_A$ under the adjoint action
\be g^{-1} \star Y_A\star g= \Lambda\fdu{A}{B}Y_B, \qquad \quad \Lambda\fdu{M}{N}=\left(
          \begin{array}{cc}
            A & B \\
            C & D \\
          \end{array}
        \right)\,.\ee
Let us look at the large twisted-adjoint rotation (\ref{largeTAdRotaion}) of the
propagator performed by $g$
\begin{align}
g^{-1}(X|Y)&\star K \exp i\left\{-y F\bar{y}+\xi y\right\}\star \tilde{g}(X|Y)=K' \exp i\left\{-y F'\bar{y}+\xi' y\right\}\,,\end{align}
\be\begin{split}\label{btobLargeTARot}
    K'&=\frac{K}{\det |A-FC|}\,, \\
    F'&=(A-FC)^{-1}(FD-B)\,, \\
    \xi'&=(A-FC)^{-1}\xi\,.
\end{split}\ee
First, propagator preserves its form\footnote{These
transformations are twistor analogs of the ones found recently for
HS fields in $d$-dimensions in \cite{Didenko:2012vh}.}. The projector property of
the propagator holds upon large twisted-adjoint rotation (\ref{largeTAdRotaion}).
This entails that $F'$ satisfies \eqref{FisSpTwo}. Second, $K'$,
$F'$ and $\xi'$ are exactly $K_X$, $F_X$ and $\xi_X$ at the point
$X$ that is defined by $g(X|Y)$. The latter implies that the two
ways of obtaining solution at point $X$, either by taking the
solution itself or first by restricting solution to $X_0$ where
$g(X_0)=1$ and then performing a large twisted adjoint rotation (\ref{largeTAdRotaion}),
give the same result. This is of course what must have been
expected.

One can invert the meaning of (\ref{btobLargeTARot}) and treat them as boundary transformations of $\rmx$ and $\eta$. Therefore boundary-to-bulk propagator is a representation of $Sp(4)$ in the bulk and on the boundary\footnote{In view of footnote \ref{ft:BiField}, there is a natural action of HS algebra on boundary
variables $\eta$ and $\rmx$ too and the propagator is an equivariant map or intertwining. }. Let us mention that similar transformations have already appeared in the context of conformal HS theories \cite{Gelfond:2008td}.

It is easy to work out the coordinate dependence of the boundary-to-bulk propagator with the help of (\ref{btobLargeTARot}) starting from the base point where $g=1$
\be
B=g^{-1}\star P_0\star \tilde{g}+c.c.+\left(\substack{\xi\leftrightarrow-\xi\\
\bar{\xi}\leftrightarrow-\bar{\xi}}\right)\,.
\ee
Indeed, taking for example $f_{\al\dal}=i\gep_{\al\dal}$, a constant $\xi^\ga$ and using
\eqref{gPoincare} one finds \eqref{PropagatorPoincare} at $\rmx=0$.

\section{Observables in higher-spin theory}\setcounter{equation}{0}\label{sec:Observables}
All physical information is encoded in master field $B(Y|X)$ even
at the nonlinear level. However, $B(Y|X)$ is not invariant under
HS transformations, moreover it transforms in the twisted-adjoint
representation of the HS algebra rather than the adjoint one,
(\ref{UnfldAB}). One way of extracting physical data is to
construct observables which are invariant under HS symmetries and
diffeomorphisms as well. This route was suggested in
\cite{Sezgin:2011hq}, elaborated further in \cite{Colombo:2010fu,
Sezgin:2012ag, Colombo:2012jx} and is closely related to the
action proposal \cite{Boulanger:2011dd,Boulanger:2012bj} for
Vasiliev equations.

That kind of observables do not strictly speaking
correspond to any conserved charges, rather to some 'initial data'
similar to Cauchy data of classical mechanics. That state of
affairs in HS theory is in many ways similar to what happens in
pure gravity, which being diffeomorphism invariant admits no
stress tensor and conserved charges. However, from the holographic
point of view, these observables are actually what one needs to
trace the correspondence. Indeed, the gauge invariant correlation
functions of the boundary theory should be rewritten in terms of
'initial data' in the bulk.

To construct observables out of $B(Y|X)$ we recall that the
$\star$-product admits uniquely defined supertrace
operation\footnote{Similar formulas with the simple trace
operation defined below appeared in \cite{Colombo:2010fu,
Colombo:2012jx}, then the Authors turned to another definition
that leads to several types of divergences that need to be regularized.}
\cite{Vasiliev:1999ba}
\be
str(F(Y))=F(0)\,,\qquad\qquad str\left(F(Y)\star
G(Y)-(-)^{\pi_G\pi_F}G(Y)\star F(Y)\right)=0\,,
\ee
where degree $\pi_F$ is defined as $F(Y)=(-)^{\pi_F}F(-Y)$. In the
bosonic HS theory that we consider the supertrace coincides with
the trace as $\pi_B=0$. Now, having a bosonic field $\Psi(Y)$ that
transforms in the adjoint
\be
\gd_{\xi}\Psi=[\Psi,\xi]_{\star }\,,
\ee
its corresponding (long)trace operator
\be
O_n=\mbox{Tr}(\Psi\star...\star\Psi)
\ee
is a HS gauge invariant quantity. The problem is that master field $B$ transforms
in the twisted-adjoint, (\ref{UnfldAB}). To fix it one observes that the
automorphism $\tilde{g}$ turns to an internal one, should we allow
$\gd$-functions in the $\star$-product algebra \cite{Didenko:2009td}
\be
\delta(y)=\int ds\, \exp i (sy)\,,
\ee
which behaves nicely under $\star$-product performing a Fourier transform
\be F(y,\bar{y})\star \delta(y)=\int ds\, F(s,\bar{y}) \exp i (sy)\,. \ee
Sandwiching  $F(y,\bar{y})$ with two $\delta(y)$'s one finds
\be\label{tilda}
\delta(y)\star F(y,\bar{y})\star \delta(y)=F(-y,\bar{y})\equiv
\tilde{F}(y,\bar{y})\,.
\ee
The seemingly asymmetric holomorphic form of $\tilde{g}$ is
fictitious since $B(-y,\bar{y})=B(y,-\bar{y})$. Therefore,
$\delta(y)$ is a map from the twisted adjoint module of HS
algebra \eqref{UnfldAB} to the adjoint one
\be
\Psi=B\star\delta(y)\,,\qquad\quad\delta_{\xi}\Psi=[\Psi,\xi]\,.
\ee
Given a set of $n$ adjoint fields $B_i(Y|X)\star\delta(y)$, $i=1...n$, their mutual HS symmetry
invariants are given by operators
\be\label{tr}
O_n=str \left(B_1\star\delta \star B_2\star\delta\star...\star
B_n\star\delta\right)\,.
\ee
\begin{wrapfigure}{l}{4cm}
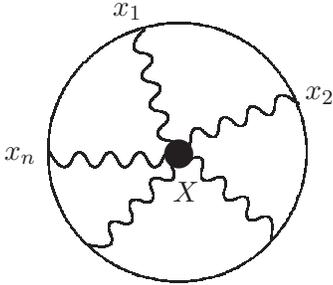

\TwistorialWitten
\label{pict}
\caption{Twistor space Witten diagram for $O_n$}
\end{wrapfigure}
The observable \eqref{tr} can be also viewed as some $n$-point
interacting vertex. The interaction is governed by the unbroken HS
symmetry via the $\star$-product operation.  In our case the set of
$B_i$ is given by boundary-to-bulk propagators
$B(X,Y|\rmx_i,\eta_i)$ from boundary points $\rmx_i$ to one and
the same bulk point $X$. To make connection with the usual AdS/CFT
paradigm, one can identify $O_n$ with an analog of the $n$-leg
Witten diagram. The interaction point
is however in the twistor space of $Y$ variables and the
interaction vertex is projected onto $Y=0$ rather than integrated
over the twistor space. No integral over the bulk position of the
vertex is taken as the interaction point drops out of $O_n$ since
a particular large gauge transformation \eqref{btobLargeTARot}
allows one to move the interaction point freely over the whole
$AdS_4$. One may think that the volume integral splits off.
It is worth emphasizing that \eqref{tr} is invariant
under all HS transformations (\ref{UnfldAB}), not only under the
$sp(4)$ subalgebra. As we see in order to reproduce the free CFT
correlators one needs to take into account only very specific
contact interactions in the twistor space\footnote{This kind of tree level $S$-matrix can be effectively obtained by substituting $\exp_\star (g B\star\delta)$ instead of $B$ into (\ref{UnfldB}) and expanding in the formal coupling $g$. }. No exchange diagrams is
needed.

Let us note that the holomorphic Fourier transform applied to
boundary-to-bulk propagator results in $\delta$-function yielding
$O_1=\mbox{derivatives}\,\,\delta(0)$. Nevertheless, as we will
soon see, all $O_n$ for $n>1$ are perfectly finite quantities. For $O_{2m}$ it is obvious. Indeed,
for even $n$ all $\delta$-function insertions can be pairwise
removed using \eqref{tilda}. If $n$ is odd one is left with a
single $\delta$-function inside the trace
\begin{align}
O_{2m}&=str(B_1\star \tilde{B}_2\star ...\star B_{2m-1}\star \tilde{B}_{2m})\,,\label{even}\\
O_{2m+1}&=str(B_1\star \tilde{B}_2\star ...\star \tilde{B}_{2m}\star B_{2m+1}\star
\delta )\,.\label{observe-odd}
\end{align}
Noticing that
\be str (F(y,\bar{y})\star \delta(y))=\int dy\, F(y,0)\,,\ee
we finally arrive at
\begin{align}
O_{2m}&=\phantom{\int dy \,}\left.(B_1\star \tilde{B}_2\star ...\star B_{2m-1}\star \tilde{B}_{2m})\right|_{Y=0}\,,\\
O_{2m+1}&=\int dy \,\left.(B_1\star \tilde{B}_2\star ...\star \tilde{B}_{2m}\star B_{2m+1})\right|_{\bar{y}=0}\,.\label{odd}
\end{align}
Despite not being immediately obvious $O_{2m}$ and $O_{2m+1}$ have the cyclic property
for $B_i(Y)$ that satisfy $B_i(Y)=B_i(-Y)$.

\paragraph{General expression.} The last step is to replace generic $B$ in $O_n$ with the boundary-to-bulk propagator. The propagator (\ref{PropagatorPoincare}) has four terms that are generated from
the first one by the action of simple discrete groups that will
survive in the final expressions. Let us define\footnote{Recall that $\theta$ is a parameter in the Vasiliev equations and it enters the propagator, see discussion after \eqref{Xidef}.}
\be\label{discr}
\rho(\xi y,\bar{\xi}\bar{y},\theta)=(\bar{\xi}\bar{y}, \xi
y,-\theta)\quad \textnormal{and}\quad \pi(\xi, \bar{\xi})=(-\xi,
-\bar{\xi})\,.
\ee
The propagator (\ref{PropagatorPoincare}) is rewritten as
\be
\qquad B=\sum_{\rho\times\pi}\Phi\equiv(1+\pi)(1+\rho)\Phi\,,\qquad\qquad\Phi(F,\xi,\theta)=K \exp i(-yF\bar{y}+\xi y+\theta)\,. \label{btobPrimary}
\ee
Then, the connected $n$-point correlation function of generating functions
of conserved currents is conjectured to be (up to a certain numerical prefactor)
\begin{align}
\langle j(\rmx_1,\eta_1)...j(\rmx_n,\eta_n)\rangle=\sum_{S_n}\sum_{\rho^n\times \pi^n} O_n\left(\Phi_1,...,\Phi_n \right)\,,\label{MostGeneral}
\end{align}
where $\Phi_i$ is a primary term (\ref{btobPrimary}) of the
propagator for the current at the $i$-th point on the boundary. It
depends on $\rmx_i,\eta_i$ through $F$, \eqref{F}, and $\xi$, \eqref{Xidef}. The sum over $\rho^n\times \pi^n$
restores antiholomorphic part of the propagators and projects it
onto the bosonic part. The sum over the symmetric group $S_n$ is
introduced to account all terms in the trace $\mbox{Tr}(B\star
\tilde{B}...B\star \tilde{B})$ except for those at coinciding
points, where the full $B$ is $\sum_i B_i$. Equivalently, one may
propose \eqref{MostGeneral} from the very beginning as the
candidate to be an observable with all the requirements being met. In the latter case
the sum over $S_n$ is required to make the expression hermitian.
In the next section we show that $O_n\left(\Phi_1,...,\Phi_n
\right)$ can be computed explicitly for arbitrary $n>1$. It is worth emphasizing
that no $z\rightarrow 0$ limit is needed. Neither should we deal with any integrals over
the bulk of $AdS$.

\section{Algebraic structure of propagators}\setcounter{equation}{0}\label{sec:btotbGroup}
The boundary-to-bulk propagators are exponents of no higher than
quadratic polynomials in $Y$ variables. While each of them has
only mixed bilinear term $y\bar y$ in the exponent, the mutual
$\star$-product of any two results in some generic bilinear exponent. These
Gaussians form a closed subalgebra under the $\star$-product. So let
us consider a generic element
\be \Phi(f,\xi, q) = \exp i{\Big( \ff{1}{2}f_{AB}Y^{A}Y^{B}+\xi^{A}Y_{A}+q\Big)}\,, \label{genericExp}\ee
and multiply two such elements within the HS algebra and see what happens.

\paragraph{$\boldsymbol{Sp(2M)\ltimes}$Heisenberg group and Cayley transform.} We may no longer restrict
ourselves to $sp(4)$ algebra, as the the following result is valid
for $sp(2M)$. First, it is easy to see that
\be \exp i (\xi Y) \star \exp i (\eta Y)=\exp i\left( (\xi+\eta)Y+\xi\eta\right)\ee
and hence elements of the form $\Phi(0,\xi,c)$ belong to the
Heisenberg group, $\Phi(0,\xi,a)\star
\Phi(0,\eta,b)=\Phi(0,\xi+\eta,a+b+\xi\eta)$,
\cite{Soloviev:2012}. It was shown in \cite{Didenko:2003aa} that
$\Phi(f_1,0,0)\star \Phi(f_2,0,0)$ computes the $Sp(2M)$-product
provided that $f_{1,2}$ are related to group elements $U_{1,2}\in
Sp(2M)$ by Cayley transform,
\be f=(1-U)(1+U)^{-1}\,.\ee
Below we consider the general case of
$\Phi(f,\xi,q)$.

Define $SpH(2M)$ as the semidirect product of $Sp(2M)$ and the
Heisenberg group, \cite{Gelfond:2008td,Gelfond:2010xs}, i.e.
$SpH(2M)$ consists of triplets $\mathcal{G}=(U_{A}{}^{B}, x_{A},
c)$, where $U_{A}{}^{B}\in Sp(2M)$ with the following product
\be
\mathcal{G}_1\diamond\mathcal{G}_2=\Big((U_1U_2)_{A}{}^{B},
x_{1A}+U_{1A}{}^{B}x_{2B},
c_1+c_2+x_{1}^{A}U_{1A}{}^{B}x_{2B}\Big)\,.\label{SpH}
\ee
The $SpH(2M)$ action can be realized by the generalized Cayley
transform $\mathcal{C}$ on the $\star$-product elements of the form
$\Phi(f,\xi,q)$, (\ref{genericExp}). For its derivation let us
first write down the $\star$-product of such two different
(\ref{genericExp}),
\begin{align}\label{StarProductGeneric}
\Phi(f_{1},\,\xi_{1},\,0)\star \Phi(f_{2},\,\xi_{2},\,0)
=\ff{1}{\sqrt{\det{|1+f_1f_2|}}}\Phi(f_{1,2},\,\xi_{1,2},\,q_{1,2})\,,
\end{align}
where
\begin{align}
&f_{1,2AB}=\ff{1}{1+f_2f_1}(f_2+1)+\ff{1}{1+f_1f_2}(f_1-1)\,,\label{f12}\\
&\xi_{1,2}^{A}=\xi_{1}^{B}\Big(\ff{1}{1+f_2f_1}{(f_2+1)}
\Big)_{B}{}^{A}+\xi_{2}^{B}\Big(\ff{1}{1+f_1f_2}{(1-f_1)}
\Big)_{B}{}^{A}\,,\label{xi12}\\
&q_{1,2}=\ff{1}{2}\Big(\ff{1}{1+f_2f_1}f_2\Big)_{AB}\xi_{1}^{A}\xi_{1}^{B}+
\ff{1}{2}\Big(\ff{1}{1+f_1f_2}f_1\Big)_{AB}\xi_{2}^{A}\xi_{2}^{B}-
\Big(\ff{1}{1+f_2f_1}\Big)_{AB}\xi_{1}^{A}\xi_{2}^{B}\label{q12}\,.
\end{align}
The generalized Cayley transform
\be
\mathcal{C}:\qquad \Phi(f, \xi, q)\to \mcG(U, x, c)
\ee
allows one to embed $SpH(2M)$ group into the $\star$-product algebra,
such that
\begin{align}
& r(\mathcal{G}_1)\,\Phi\Big(
f(\mathcal{G}_1),\,\xi(\mathcal{G}_1),\,q(\mathcal{G}_1)\Big)\star
r(\mathcal{G}_2)\,\Phi\Big(
f(\mathcal{G}_2),\,\xi(\mathcal{G}_2),\,q(\mathcal{G}_2)\Big)=\\
&\qquad\quad r(\mathcal{G}_1\diamond\mathcal{G}_2)\,\Phi\Big(
f(\mathcal{G}_1\diamond\mathcal{G}_2),\,\xi
(\mathcal{G}_1\diamond\mathcal{G}_2),\,q(\mathcal{G}_1\diamond\mathcal{G}_2)\Big)\,.
\end{align}
Its explicit form reads
\begin{align}
&f_{AB}(\mathcal{G})=\Big(\ff{U-1}{U+1}\Big)_{AB}\,,\label{f(U)}\\
&r(\mathcal{G})=\ff{2^{M/2}}{\sqrt{\det{|1+U|}}}\,,\\
&\xi_{A}(\mathcal{G})=\pm 2\Big(\ff{1}{1+U}\Big)_{A}{}^{B}x_B\,,\\
&q(\mathcal{G})=c+\ff{1}{2}\Big(\ff{U-1}{U+1}\Big)_{AB}x^Ax^B\,.
\end{align}
In principle, the $SpH$ structure of propagators allows one to
compute the super-trace that gives HS observables according to the
following recipe: (i) map boundary-to-bulk propagators
$\Phi_1$,...,$\Phi_n$ into elements $\mcG_1$,..., $\mcG_n$ of
$SpH$ by the inverse Cayley transform; (ii) compute the product
$\mcG=\mcG_1\diamond...\diamond\mcG_n$; (iii) take the central
part of $\mcG$ and map it back. This method fails, however,
when \eqref{genericExp} is such that $f^2=I$.

\paragraph{Peculiarity of propagators.} As we have already mentioned, boundary-to-bulk
propagators (\ref{btobPrimary}) are not generic elements because
the quadratic form $f$ is involutary $f^2=I$,
\be f=\left(
        \begin{array}{cc}
          0 & -F \\
          -F^T & 0 \\
        \end{array}
      \right),\quad F\in Sp(2) \quad\Longrightarrow\quad\det f=1\,,\qquad f^2=I \label{fdefinition}\,.
\ee
Its Cayley transform cannot be inverted to get the group element.
The $\star$-product is still well-defined. Matrices $f$ that are
close to being a square root of a unit matrix give group elements
that are close to infinity in a sense of having very large matrix
elements. General formulae (\ref{f12})-(\ref{q12}) being examined
on the space of $\Phi(f,\xi,q)$, $f^2=I$, reduce to
\begin{align}
&f_{1,2}=f_1\circ f_2=\ff{1}{f_1+f_2}(2+f_2-f_1)\,,\label{fr12}\\
&\xi_{1,2}^{A}=\ff12\xi_{1}^{B}(1+f_1\circ f_2)
_{B}{}^{A}+\ff12\xi_{2}^{B}(1-f_1\circ f_2)_{B}{}^{A}\,,\label{xir12}\\
&q_{1,2}=\ff{1}{8}\{f_1, f_2\}_{\circ
AB}(\xi_{1}^{A}\xi_{1}^{B}+\xi_{2}^{A}\xi_{2}^{B})-
\ff12(1+\ff12[f_1,f_2]_{\circ})_{AB}\xi_{1}^{A}\xi_{2}^{B}\label{qr12}\,.
\end{align}

We can view (\ref{fr12})-(\ref{qr12}) as the extension of the
group action to certain cell attached to $SpH$ at infinity. The
small-cells that are related by inversion combined with reflection cannot be attached to
$SpH$ simultaneously, for $f(-\rmx/\rmx^2)=-f$ and the denominator of (\ref{fr12}) vanishes.

In order to compute correlation functions one can proceed in two
ways at least. Either regularize boundary-to-bulk propagators to
shift them to the big cell that is to make them generic, then
compute using $SpH$, (\ref{SpH}), and remove the regularization at
the end. For example, one can duck the issue by redefining $f_i\to
\gep f_i$, where $\gep$ is an arbitrary number $\gep\neq 1$. This
makes Cayley transform well-defined and allows one to extract
trace \eqref{even} as the limit of the inverse Cayley transform at
$\gep\to 1$. In practice, the limit $\gep\to 1$ turns out to be
difficult to compute and we found it simpler, to apply the
small-cell rules (\ref{fr12})-(\ref{qr12}) directly, leaving aside
any regularization problems.

Since $f_{1,2}^2=I$, eq.
(\ref{fr12}) defines a product on the space of square roots of
unity. Its straightforward properties are
\begin{align}
&(f_{1}\circ f_{2})_{AB}=(f_1\circ f_2)_{BA}\label{sym}\,,\\
&f_1\circ(f_2\circ f_3)=(f_1\circ f_2)\circ f_3\label{ass}\,,\\
&f\circ f=f\label{proj}\,,\\
&(-f_1)\circ(-f_2)=-f_2\circ f_1\,,\label{sign}\\
&f_1\circ f_2\circ f_3=f_1\circ f_3\label{loc}\,.
\end{align}
Associativity \eqref{ass} is induced by the associativity of the
$\star$-product. Last property \eqref{loc} is due to the uniqueness
of the $\star$-product element $F=\exp{(\ff12(f_{-}\circ
f_{+})_{AB}Y^{A}Y^{B})}$ which is annihilated by $Y^{\pm}=(1\pm
f_{\pm})Y$ from the right and left for $Y^+$ and $Y^-$,
respectively. This last 'forgetful' property \eqref{loc} is very important and it
will imply that only fields that are adjacent along the $n$-cycle
will effect inside \eqref{MostGeneral}.

Let us note that despite seemingly $Sp(2)$-origin of the
propagator, i.e. a single $\Phi$ contains just $y F\bar{y}$ and
not a generic $Y\!AY$ with $A^2=I$, the product $\Phi_1\star
...\star \Phi_n$ falls into $Sp(4)$. Matrices of the particular
form \eqref{fdefinition} belong to $Sp(2)\times$split-complex
numbers.

In practice it is convenient to rewrite \eqref{fr12}-\eqref{qr12}
in terms of the following projectors
\be
\pi^{\pm}_{ij}=\ff12(1\pm f_i\circ f_j)\,,\qquad
\pi_{ij}^{\pm}\pi_{ij}^{\pm}=\pi_{ij}^{\pm}\,,\qquad \pi_{ij}^{\pm}\pi_{ij}^{\mp}=0\,,
\ee
which have the properties
\be
\pi^{+}_{ij}\pi^{+}_{ik}=\pi^{+}_{ik}\,,\qquad
\pi^{-}_{ij}\pi^{+}_{ik}=0\,.
\ee
Using these projectors and \eqref{sym}-\eqref{loc} makes the
calculation of the exponent of $\textnormal{Tr}(\Phi_1\star
...\star \Phi_n)$ rather straightforward resulting in,
\be\label{exp}
\textnormal{Tr}(\Phi_1\star ...\star \Phi_n)\sim \exp
i\Big(\ff18\sum_{j=1}^{n}(f_j\circ f_{j-1}+f_{j+1}\circ
f_j)_{AB}\xi^{A}_{j}\xi^{B}_{j}+\ff14\sum_{j=1}^{n}(1+f_{j+1}\circ
f_j)_{AB}\xi^{A}_{j+1}\xi^{B}_{j}\Big)\,,
\ee
where the sum is
understood over $1...n\mod n$.
One can see that only adjacent points contribute to
the final result (the sum goes along the cycle) which is valid for
$sp(2M)$ case as well. For $n=2$, exponent \eqref{exp} is given by
\eqref{qr12}, which appears to be exactly the same provided that the specific form of $\xi_A=(\xi_\ga,0)$, $\bar{\xi}_A=(0,\bar{\xi}_{\gad})$ is taken into account. Calculation of prefactor in \eqref{exp} is much
trickier and is left for Appendix B. The prefactor reads
\be\label{det}
\ff{1}{2^{M(n-1)}\sqrt[4]{\prod\limits_{i=1}^{n}\det{|f_i+f_{i+1}|}}}
\ee
provided the following condition is met
\be\label{cond}
\det{|f_i\circ f_j+f_k\circ f_l|}=\det{|f_j\circ f_i+f_l\circ
f_k|}
\ee
being the case for propagators. Using \eqref{sign} we see that eq.
\eqref{cond} makes the determinant invariant upon simultaneous
sign flip $f_i\to -f_i$ for all $i$.

When $n$ is odd the additional integration is needed \eqref{odd}
in extracting correlation functions. This holomorphic integration
yields a prefactor to the determinants \eqref{det}
\be\label{delta-det}
\ff{1}{\sqrt{\det(f_1\circ f_n)_{\al\gb}}}
\ee
and a contribution to the exponent \eqref{exp} of the form
\be\label{delta-exp}
-\ff{i}{2}(f_1\circ
f_n)^{-1}_{\al\gb}\xi_{1,n}^{\al}\xi_{1,n}^{\gb}\,.
\ee
As we will see, in both even and odd cases the arising structures
are simply the conformal ones and the final result does not
distinguish even and odd cases very much.

\section{$N$-point functions}\setcounter{equation}{0}\label{sec:Npoints}
\paragraph{Conformal structures.} Given a correlator of tensor operators $j(\rmx_a,\eta_a)$
at points $\rmx_a$, whose tensor structure is encoded in terms of
polarization spinors $\eta_a$, it may depend on few conformally
invariant quantities that involve $\eta$'s. This is on top of the
ambiguity in functions of conformally invariant ratios
$x_{12}x_{34}/(x_{14}x_{24})$.

There are two basic structures \cite{Costa:2011mg, Giombi:2011rz}
that may appear inside correlation functions. One structure
depends on two points and two polarization spinors,
$P_{ab}=-P_{ba}$, another one depends on three points and one
polarization spinor (quadratic in it), $Q_{bc}^a=-Q^a_{cb}$, where
indices $a,b,...$ enumerate boundary sources.

We will present $P_{ab}$ and $Q_{bc}^a$ together with their 'bulk
realization'. To do so\footnote{Since the observables do not depend on coordinate choice in the bulk we have set $x=0$, $z=1$ for simplicity reason.} let us define $a$-$b$ points intertwinings
$F_{ab}$
\begin{align}
F_{ab}^{\ga\gb}&=F_a^{\ga\gad} F_b{}\fud{\gb}{\gad}\,, \qquad\qquad\qquad \det{ (\epsilon-F_{ab})}=\frac{4 \rmx_{ab}^2}{(1+\rmx_a^2)(1+\rmx_b^2)}\,,\\
F_{ab}&{}\fdu{\ga}{\gb}F_{ab}{}^{\ga\gc}=\epsilon^{\gb\gc}\,, \qquad\qquad\qquad F_{ab}{}\fud{\ga}{\gb}F_{ab}{}^{\gc\gb}=\epsilon^{\ga\gc}\,,\end{align}
then the conformal structures can be found to have the form
\begin{align}
P_{ab}&=2\xi_b(\epsilon-F_{ab})^{-1}\xi_a\sim \eta_b \rmx^{-1}_{ab}\eta_a\,,\\
Q^a_{bc}&=4\xi_a(\epsilon-F_{ca})^{-1}(\epsilon-F_{bc})(\epsilon-F_{ab})^{-1}\xi_a\sim\eta_a(\rmx^{-1}_{ab}+\rmx^{-1}_{ca})\eta_a\,.
\end{align}
These structures can be identified as building blocks of simplest correlators
\begin{align}
\langle j_{s_1}(\rmx_1,\eta_1) j_{s_2}(\rmx_2,\eta_2) \rangle&\sim\frac{1}{\rmx^2_{12}} \delta_{s_1,s_2}(P_{12})^{s_1+s_2}\,,\\
\langle j_{s_1}(\rmx_1,\eta_1) j_0(\rmx_2) j_0(\rmx_3)\rangle&\sim\frac{1}{\rmx_{12}\rmx_{23}\rmx_{31}} (Q^1_{23})^{2s_1}\,.
\end{align}
Let us note that one and the same conformal structure can be
rewritten in several different ways in the bulk since both $\xi$
and $\bar{\xi}$ originate from the same boundary spinor $\eta$.
Thus defined conformal structures can be rewritten in terms of
$\circ$-product as follows
\be\label{confstr}
Q^{a}_{bc}=\ff18\xi_a \big(f_a\circ (-f_b)+(-f_c)\circ
f_a\big)\xi_a\,,\qquad P_{ab}=-\ff14\xi_a\big(1+(-f_b)\circ
f_a\big)\xi_b\,,
\ee
where the origin of minus signs can be understood from alternation
in \eqref{even}. These are the building blocks of \eqref{exp}.
Note, that \be\label{QPbars} Q(\xi)=Q(\bar\xi),\qquad P(\xi_a,\xi_b)=P(\bar\xi_a,\xi_b)=-P(\xi_a,\bar\xi_b)=-P(\bar\xi_a,\bar\xi_b)\,.\ee

\paragraph{$2$-point functions} Let us discuss first the simplest
case of two-point functions without any reference to the
underlying projector structure of propagators. We first find
\begin{align*}
(12)\equiv\Phi(F_1,\xi_1)\star \Phi(F_2,\xi_2)&=\frac{K_1 K_2}{|1+F_{12}|}\exp i\left(-(F_2\bar{y}+y+\xi_2)(1+F_{12})^{-1}(F_1\bar{y}-y+\xi_1)\right)\,,\\
(1\bar{2})\equiv\Phi(F_1,\xi_1)\star \Phi(F_2,\bar{\xi}_2)&=\frac{K_1 K_2}{|1+F_{12}|}\exp i\left(-(F_2\bar{y}+y)(1+F_{12})^{-1}(F_1\bar{y}-y+\xi_1+F_1\bar{\xi})+\bar{\xi}_2\bar{y}\right)
\end{align*}
and a similar expressions for $\Phi(F_1,\bar{\xi}_1)\star \Phi(F_2,\bar{\xi}_2)$ and  $\Phi(F_1,\bar{\xi}_1)\star \Phi(F_2,{\xi}_2)$. Next we change $F_2\rightarrow -F_2$ and
$\xi_2\rightarrow -\xi_2$ and set $Y=0$ to get $str(\Phi\star \tilde{\Phi})$
\begin{align}
(12)&=(\bar{1}\bar{2})=\frac{1}{4\rmx_{12}^2}\exp i 2P_{12}\,, &
(1\bar{2})&=(\bar{1}{2})=\frac{1}{4\rmx_{12}^2}\,.
\end{align}
Finally we need to sum up all contributions and project onto the
bosonic part, (\ref{MostGeneral}), resulting in
\be \langle j(\rmx_1,\eta_1) j(\rmx_2,\eta_2)\rangle=\frac{4}{\rmx_{12}^2} (1+\cos 2\theta\, \cos 2P_{12})\,,\ee
where we recall that $\theta=0$ and $\theta=\pi/2$ for free
boundary bosons and fermions, respectively. Note that $\Delta(j_0)=2$ in the free fermion vector model and
$\langle j_0 j_0\rangle$ is not reproduced by $\Delta=1$ propagator.

Let us note that as usual within the AdS/CFT, two-point function
can be extracted already from the boundary-to-bulk propagator.
Indeed, taking $z\rightarrow0$ one naively finds for the primary
term of (\ref{PropagatorPoincare}), that the
singleton-antisingleton vacuum $\exp{y_\ga \bar{y}^\ga}$
suppresses all other contributions,
\be B\rightarrow z \exp{y_\ga \bar{y}^\ga}+O(z^2)\,.\ee
As it is explained in \cite{Vasiliev:2012vf}, equation
(\ref{UnfldB}), whose solution  the propagator is, has a
meaningful limit $z\rightarrow0$ at the conformal boundary, which
gives equations for conserved currents. The limiting procedure for
the solutions is
\be B\rightarrow z \exp{y_\ga \bar{y}^\ga}\, T(yz^{\frac12},\bar{y}z^{\frac12})\,.\ee
Extracting $T$ according to this prescription one finds
\be T=\exp y(x-\rmx)^{-1}\eta\,,\ee
which is a correct two-point function of conserved currents, where $y$ plays the role of polarization spinor at point $x$. Then it is obvious that (\ref{btobConservation}) holds true in the boundary limit.

\paragraph{Example: 3-point.} We present only the primary
term of (\ref{MostGeneral}), the rest can be generated by the
action of the symmetric group $S_3$ and involutions $\rho$ and $\pi$,  whose
action is described below. Using \eqref{exp}, \eqref{det} and
\eqref{delta-det}, \eqref{delta-exp} we obtain
\begin{align}\label{threepoint}
(123)=\frac{1}{16\,\rmx_{12} \rmx_{23}\rmx_{31}}\exp
i\left\{(Q^1_{32}+Q^2_{13}+Q^3_{21})+(P_{12}+P_{23}-P_{31})+3\theta\right\}\,,
\end{align}
where we also used that $M=2$ for $Sp(4)$ and the explicit form of
$f_i$ \eqref{fdefinition} which gives
\be
\sqrt{\det{|f_i-f_{j}|}}=4 \rmx^2_{ij} K_i K_j\,.
\ee
\paragraph{Example: 4-point.} Analogous computation gives
\begin{align}\label{fourpoint}
(1234)=\frac{1}{64\,\rmx_{12} \rmx_{23}\rmx_{34}\rmx_{41}}\exp i\left\{(Q^1_{42}+Q^2_{13}+Q^3_{24}+Q^4_{31})+(P_{12}+P_{23}+P_{34}-P_{41})+4\theta\right\}\,.
\end{align}

\paragraph{$\boldsymbol{n}$-point.}
The calculation of the $n$-point function is different for even and
odd $n$ due to the difference in  observables \eqref{even} and
\eqref{observe-odd}. This difference forces one to take into
account additional terms \eqref{delta-det} and \eqref{delta-exp}
in odd case. The final result, however, has the unique closed
form. To formulate the final result it is convenient to define
\begin{align}\label{PQQ}
& P_i\equiv P_{i,i+1}(-)^{\delta_{i,n}}\,,&& Q_i\equiv
Q^i_{i-1,i+1}\,, && Q=\sum_i Q_i\,,\end{align} where $(-)^{\delta_{j,n}}$ accounts
for the sign flip of the last term, c.f. \eqref{threepoint},
\eqref{fourpoint}. Then the primary $(12...n)$ term is (the sums
and the product are understood over $1...n\mod n$),
\begin{align}
O_n(\Phi_1,...,\Phi_n)=(12...n)=\frac{1}{2^{2n-2}\prod_i|\rmx_i-\rmx_{i+1}|}\exp i\left\{\sum_j Q_j+\sum_k P_k +n\theta\right\}\,.
\end{align}
Formally $2-pt$ also fits into this formula.

To sum up to the correlation function according to \eqref{MostGeneral} we have to know the action of
$\rho_i$ and $\pi_i$ \eqref{discr}. $Q$-structure is left
unaffected by $\rho$ and $\pi$, while $\rho_i$ changes the sign of
$P_{i-1}$ and $\theta$, and $\pi_i$ flips the sign of $P_i$ and
$P_{i-1}$, which follows from definitions \eqref{confstr} and properties \eqref{QPbars}. The sum over $\rho^n$ gives
\begin{align}
\sum_{\rho^n}O_n(\Phi_1,...,\Phi_n)=\frac{1}{2^{n-2}\prod_i|\rmx_i-\rmx_{i+1}|}\exp i\left\{Q\right\}\prod_k \cos \left(P_k +\theta\right)\,.
\end{align}
The sum over $\pi^n$ affects $P_k$'s only and can be evaluated using
\be \sum_{\pi^n}\prod_k\cos \left(P_k +\theta\right)=2^n \cos^n \theta \prod_k\cos \left(P_k\right)+
2^n\sin^n \theta\prod_k\sin \left(P_k\right) \,. \ee To sum up
over $S_n$ it is convenient to introduce the dihedral group, the
symmetry group of a regular $n$-gon, $D_n$. It contains $2n$
elements, $n$ rotations and $n$ reflections. Elements of $D_n$ map
$Q$, $P$ structures and the prefactor to themselves, while those
of $S_n/D_n$ generate new permutations. For example $S_4/D_4$
produces three terms of the $4$-point functions of the scalar
operators $j(\rmx_i,0)$
\be \langle j(\rmx_1,0)...j(\rmx_4,0)\rangle\sim
(\rmx_{12}\rmx_{23}\rmx_{34}\rmx_{41})^{-1}+
(\rmx_{13}\rmx_{32}\rmx_{24}\rmx_{41})^{-1}+
(\rmx_{12}\rmx_{24}\rmx_{43}\rmx_{31})^{-1}\,.\ee It is easy to see
that the rotations $r$ of $D_n$ do not affect $Q$, \eqref{PQQ},
$r(Q)=Q$ while reflections $s$ flip the sign $s(Q)=-Q$. The same
time for $P=\prod_k f_\sigma(P_k)$, where
\be f_n(x)=\left\{
     \begin{array}{ll}
      \cos x, &  n\quad \hbox{even;} \\
       \sin x, & n\quad \hbox{odd\,,}
     \end{array}
   \right.
\ee
and $f_n$ possesses a
well-defined parity $\sigma$, $f_\sigma(-x)=(-)^\sigma
f_\sigma(x)$, one finds $r(P)=P$ for rotations and
$s(P)=(-)^{n\sigma}P$ for reflections. Therefore the sum over
$D_n$ affects $Q$'s producing $f_{n\sigma}(Q)$. It is useful to
illustrate the reasoning above with the picture on which each three adjacent
points of the $n$-gon correspond to the $Q$ structure and each two to the $P$ structure. Evidently
rotations do not affect $P$ and $Q$, and reflections produce a sign factor,
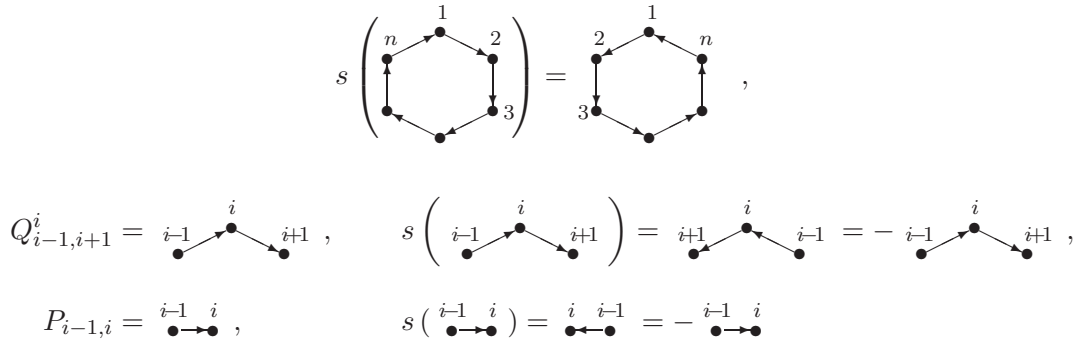
\begin{align*} &s\left(\NgonA\right)=\NgonB\quad\,, \end{align*}
\begin{align*} Q^i_{i-1,i+1}&=\QstrA\,,&& s\left(\QstrA\right)=\QstrB=-\QstrA\,,\\
\rule{0pt}{20pt}P_{i-1,i}&=\PstrA\,,&& s\left(\PstrA\right)=\PstrB=-\PstrA \end{align*}
Finally, summing up all contributions we get the generating function of connected correlators
\begin{align}
\langle &j(\rmx_1,\eta_1)...j(\rmx_n,\eta_n)\rangle=\nn \\
&\sum_{S_n}\frac{4}{\prod_i|x_i-x_{i+1}|}\left(
\cos\left(Q\right)\cos^n\theta\prod_k  \cos \left(P_k \right)
+f_{n}\left(Q\right)\sin^n\theta\prod_k \sin \left(P_k
\right)\right)\,, \label{npointAnswer}
\end{align}
where $\theta=0$ ($\theta=\pi/2$) for bosonic (fermionic) vector-model, respectively. The result is real and totally symmetric over $n$ legs as it must be. Note, that sum \eqref{npointAnswer} is over $S_n$ and not $S_n/D_n$ at the price of counting the same terms $2n$ times. For
example, one finds a familiar expressions, \cite{Giombi:2010vg,
Colombo:2012jx}, for the $3$-point functions
\begin{align} \label{threepointAnswer}
\langle j(\rmx_1,\eta_1)...j(\rmx_3,\eta_3)\rangle_{boson}&=
\frac4{\rmx_{12}\rmx_{23}\rmx_{31}}\cos(Q^1_{32}+Q^2_{13}+Q^3_{21})\cos(P_{12})\cos(P_{23})\cos(P_{31})\\
\langle j(\rmx_1,\eta_1)...j(\rmx_3,\eta_3)\rangle_{fermion}&=
\frac4{\rmx_{12}\rmx_{23}\rmx_{31}}\sin(Q^1_{32}+Q^2_{13}+Q^3_{21})\sin(P_{12})\sin(P_{23})\sin(P_{31})
\end{align}
The 4-point function result is
\begin{align}
\langle
j(\rmx_1,\eta_1)...j(\rmx_4,\eta_4)\rangle_{boson}=&
\ff{4}{\rmx_{12}\rmx_{23}\rmx_{34}\rmx_{41}}
\cos{(Q^1_{42}+Q^{2}_{13}+Q^{3}_{24}+Q^{4}_{31})}\times\nn\\
&\times
\cos{P_{12}}\cos{P_{23}}\cos{P_{34}}\cos{P_{41}}+(1\leftrightarrow
4)+(1\leftrightarrow 2)\\
\langle
j(\rmx_1,\eta_1)...j(\rmx_4,\eta_4)\rangle_{fermion}=&
\ff{4}{\rmx_{12}\rmx_{23}\rmx_{34}\rmx_{41}}
\cos{(Q^1_{42}+Q^{2}_{13}+Q^{3}_{24}+Q^{4}_{31})}\times\nn\\
&\times
\sin{P_{12}}\sin{P_{23}}\sin{P_{34}}\sin{P_{14}}+(1\leftrightarrow
4)+(1\leftrightarrow 2)
\end{align}
and the case $\langle j_2 j_0 j_0 j_0\rangle_{boson}$ matches $(6.9)$ of \cite{Maldacena:2011jn}
provided the disconnected part is excluded.
The generating function of correlators we found does not contain $j_0$ in the case of the free fermion theory, since it has weight $2$ and is not covered by the $\Delta=s+1$ propagator we used.

The correlation functions we found have a factorized form
\be
\langle j...j\rangle_\theta = \langle j...j\rangle_{boson} \cos^n
\theta+\langle j...j\rangle_{fermion}\sin^n \theta
\ee
Let us note that only for $\theta=0,\pi/2$ the boundary fall-off
of the propagator is preserved by the HS symmetry,
\cite{Vasiliev:2012vf}. Therefore, our results are meaningful only
for $\theta=0,\pi/2$ and there is no contradiction with the
implications of a slightly broken HS symmetry that imply for
$3$-point functions \cite{Maldacena:2012sf}
\be \langle jjj\rangle_\theta = \langle jjj\rangle_{boson} \cos^2 \theta+\langle jjj\rangle_{fermion}\sin^2 \theta+\langle jjj\rangle_{odd} \cos \theta \sin \theta\label{SlightlyBroken}\,.\ee
For generic $\theta$ \eqref{npointAnswer} is just a generating function for correlation functions of conserved currents.

\section{Conclusions}\setcounter{equation}{0}\label{sec:Conclusion}
We found all correlation functions of the CFT that is dual to the
Vasiliev HS theory in four dimensions with boundary conditions that do not break HS
symmetries at all orders in perturbation theory. These are
correlation functions of conserved currents of the free $O(N)$
vector model, either bosonic or fermionic one,
\cite{Maldacena:2011jn}.

\begin{wrapfigure}{l}{7cm}
\noindent\begin{tabular}{|c|c|}
  \hline CFT & \centering HS\tabularnewline\hline
 \rule{0pt}{14pt} $\langle j...j\rangle$ & $tr(\Phi\star...\star\Phi)$\tabularnewline
 \rule{0pt}{14pt} $[Q,j]=\sum \pl..\pl j$ & $\delta \Phi=[\Phi,\xi]$\tabularnewline
 \rule{0pt}{14pt} $Q\langle j...j\rangle=0$& $\delta\, tr(\Phi\star...\star\Phi)\equiv0$ \tabularnewline
  \hline
\end{tabular}
\end{wrapfigure}
In drawing the parallels between the CFT and HS theory languages
let us mention the following. Provided that $\xi$ is a propagator
itself the HS gauge transformations $\delta \Phi=[\Phi,\xi]$  are
analogous to the action $[Q,j]=\sum \pl..\pl j$ of HS charges on
the currents, which were heavily used in \cite{Maldacena:2011jn}.
The vacuum expectation value $\langle j...j\rangle$ is equivalent
to taking the trace $Tr(\Phi\star...\star\Phi)$. The Ward
identities are equivalent to the invariance of the trace under the
adjoint transformations.

That the result has a simple form \eqref{npointAnswer} of exponent of a linear combination of $P$, $Q$
conformally invariant structures is a consequence of the fact that the propagator is a Gaussian in generating oscillators $Y$ and
the $\star$-product of two Gaussians is a Gaussian again and hence it is no more than quadratic in polarization
spinors thus being linear in $P$'s and $Q$'s.

The observables $O_n$ can also be viewed as $n$-point contact
Witten diagrams, see discussion after \eqref{tr}. Therefore, $O_n$
could be understood as the interaction vertices that are (i) fully
determined by unbroken HS symmetry; (ii) sufficient to recover all
correlators of the dual CFT.

The exact HS symmetry turns out to be powerful enough to restrict
all correlation functions, \cite{Maldacena:2011jn}. It even allows
one to find all correlation functions in a closed form as
invariants of the HS symmetry. Therefore, HS algebra is a relevant
replacement for the Virasoro one when trying to find exactly
solvable CFT's in higher-dimensions. The important difference is that all CFT's with exact HS symmetry are free ones.

It opens the avenue to the study of broken HS symmetry, which may
still render the model to remain solvable in some sense. Another
choice of boundary conditions makes higher-spin symmetries broken
by $O(1/N)$ effects. Promisingly, the higher-spin symmetry still
restricts the form of correlation functions,
\cite{Maldacena:2012sf}. The CFT$_3$ $3$-point functions may
contain three different structures, \eqref{SlightlyBroken}, thus
giving a freedom for two relative coefficients. These two turn out
to be not independent, leaving only one free parameter which can
be identified with the parameter $\theta$ in Vasiliev equations.
It is interesting to trace explicitly the way the HS symmetry
restricts correlation functions of CFT's that slightly break HS
symmetries.

It would be instructive to extend the results of this paper to
various free CFT's: (i) free bosons in $d$-dimensions, where
boundary-to-bulk propagators for Vasiliev master fields have been
recently found in \cite{Didenko:2012vh}; (ii) free fermions in
$d$; (iii) free $\mathcal{N}=4$ SYM, where the projector structure
of boundary to bulk propagators could be quite interesting and the
implications of HS symmetry when the interactions are turned on
remain to be seen; (iv) higher-spin singletons or (anti)-self dual
fields in $d=2n$ dimensions \cite{Bekaert:2009fg}.

At present we see three ways of computing correlation functions in the Vasiliev theory:

{\bf 1.} One may use the $S$-matrix approach solving bulk equations
to the $n$-th order where the solution $B_{n}(x,z)$
\be
B_{n}(x,z)=\int K_n\left[ B(\rmx_1),... ,B(\rmx_n)\right]
\ee
is given by some complicated integral kernel $K_n$ acting on
the product of initial data $B(\rmx_i)$ with sources at boundary
points $\rmx_i$. At the end $(n+1)$-correlation functions are
extracted by taking $z\rightarrow 0$ limit of the solution. This
program was performed for $3$-point functions in
\cite{Giombi:2009wh, Giombi:2010vg}.

{\bf2.} One may compute fair observables \cite{Colombo:2010fu, Colombo:2012jx}, i.e. insert solutions $B_{n}(x,z)$
up to the $n$-th order into observables $O(B,...,B)$, which again calls for a good knowledge of $K_n$. This completion by nonlinear corrections is a part of the general framework for HS quantum theory of \cite{Sezgin:2011hq,Colombo:2010fu, Sezgin:2012ag, Colombo:2012jx,Boulanger:2011dd,Boulanger:2012bj}

{\bf 3.} One may compute observables \eqref{even},
\eqref{observe-odd} to the leading order, which are fully and
explicitly governed by the HS symmetry. This is the simplest of
the three and allows one to find all correlation functions
(\ref{npointAnswer}) when higher-spin symmetry is unbroken.

Recent result of \cite{Vasiliev:2012vf} indicates that for the
boundary conditions conjecturally corresponding to free CFT's the HS
symmetry seems to remain unbroken to all orders, which implies,
technical detail to be yet clarified though, that No.1 and No.2
are quite long ways to arrive at No.3, which is in accordance with
\cite{Maldacena:2011jn}. This is also in accordance with
\cite{Colombo:2012jx}, where the first correction to the
observables due to HS interactions in the bulk has been found to
coincide with the leading term. The correlators found in this
paper, (\ref{npointAnswer}), are point-split and require certain
contact terms to be added at coincident points. As was noted in
\cite{Colombo:2012jx} the HS theory may also deliver such contact
terms.

The bulk point which is connected to the boundary current and the
bulk coordinates themselves can be easily changed by a large
twisted-adjoint rotation (\ref{largeTAdRotaion}). The observables
we calculated are manifestly invariant under such transformation.
Therefore, once certain general and coordinate-invariant
properties of the propagators are understood, there is no need to
make any reference to $AdS$. In particular one may try to take the
boundary limit $z\rightarrow0$ for propagators first and then
compute the observables 'on the boundary'.

Interestingly, most of the calculations in the present paper do
not rely on space-time dimension being four and  are formally
valid for any $Sp(2M)$, viewing $Sp(2M)$ as a generalized
conformal symmetry,
\cite{Vasiliev:2001dc,Vasiliev:2002fs,Gelfond:2006be,
Gelfond:2008ur,Gelfond:2008td,Gelfond:2010xs,Gelfond:2010pm}. On
this symplectic way our work generalizes \cite{Vasiliev:2003jc},
where two and three point correlation functions of $Sp(2M)$-scalar
and vector fields were found.

Another possible application of the elaborated technique can be
the calculation of $n$-point interaction vertices and Neumann
coefficients in string field theory along the lines of Moyal
formulation \cite{Bars:2001ag, Bars:2002yj}.

\section*{Aknowledgements}
\label{sec:Aknowledgements}
We would like to thank Kostya Alkalaev, Nicolas Boulanger, Stefan Fredenhagen,
Jianwei Mei, Ilarion Melnikov, Oleg Shaynkman, Per Sundell and especially Olga Gelfond, Massimo Taronna,
Mikhail Vasiliev and Alexander Zhiboedov for valuable discussions and comments. The work of E.S. was
supported by the Alexander von Humboldt Foundation. The work of
V.D. was supported in part by the grant of the Dynasty Foundation.
The work of E.S. and V.D. was supported in part by RFBR grant
No.11-02-00814, 12-02-31837 and Russian President grant No. 5638.

\begin{appendix}
\renewcommand{\theequation}{\Alph{section}.\arabic{equation}}

\section{Boundary-to-bulk propagators}
\setcounter{equation}{0}\setcounter{section}{1}\label{sec:btob}
The right choice of the ansatz for the propagator proceeds from:
(i) $B(Y=0)$ is just a weight $\Delta=1$ scalar field, propagator
for which is well-known; (ii) the derivatives $D_{\ga\gad}$ of the
fields correspond to expansion in translation generator
$T_{\ga\gad}F^{\ga\gad}$ with yet unknown $F^{\ga\gad}$; (ii) the
boundary polarization spinor $\eta$ must be transported to the
bulk spinor $\xi$ by certain parallel transport bispinor
$\xi^\ga=\Pi^{\ga\gb}\eta_\gb$, {\em{idem.}} for $\bar{\xi}$. This
suggests the ansatz to be
\begin{align}
u&=T_{\ga\gad}F^{\ga\gad}\,, & v&=i\xi^\ga y_\ga\,, &
\bar{v}&=i\bar\xi^{\gad} \bar y_{\gad}\,, &
B&=K\,f(u,v)+K\,f(u,\bar{v})\,.
\end{align}
Eq. (\ref{UnfldB}) leads to the following set of equations
($h^{\ga\gad}=\Omega^{\ga\gad}$ is a vierbein)
\begin{align}
&d \ln{K}+\frac12 F_{\ga\gad} h^{\ga\gad}\pl_u \ln{f}=0
&&(DF^{\ga\gad}\pl_u+2h^{\ga\gad}+\frac12 F\fud{\ga}{\gdd}h^{\gd\gdd}F\fdu{\gd}{\gad}\pl^2_u)f=0\,,\label{AppbtobB}\\
&\left(D\xi^\ga+\frac12 F\fud{\ga}{\gdd} \xi_\gd h^{\gd\gdd}\pl_u\right)\pl_vf=0\,,
&&\left(D\xi^{\gad}+\frac12 F\fdu{\gd}{\gad} \xi_{\gdd} h^{\gd\gdd}\pl_u\right)\pl_{\bar{v}}f=0\,.\label{AppbtobD}
\end{align}
(\ref{AppbtobB}) determines $F$ up to some factor and gives
$f=e^{-2u}h(v)$. The rest of the equations can be easily solved by
looking only at the components along $dz$. In particular it is
obvious that the dependence on $v$ can be arbitrary, which just
encodes a freedom in normalization of propagators of all spins
independently. It is convenient to choose $h(v)=e^v$.
\section{Determinants}
\setcounter{equation}{0}\setcounter{section}{2}\label{sec:dets} To
calculate prefactor \eqref{det}, it is convenient to consider even
$n\to 2n$. The result for odd $n$ is reproduced from even case by
setting say $f_2=f_1$. Indeed, we can set all polarizations to
zero, for they do not effect the determinants, and all $\Phi_i$
are projectors $\Phi_i\star \Phi_i\sim \Phi_i$.

The determinants arising from $\star$-product can be rewritten in
terms of $\circ$ -product as follows
\be
\det{|\ff{1}{1+f_1f_2}|}=\det{|\ff{1}{f_1+f_2}|}=\ff{1}{2^{4M}}\det{|f_1\circ
f_2+f_2\circ f_1|}
\ee
Applying this identity for $\det{|f_1\circ f_2+f_3\circ f_4|}$ and
using \eqref{loc} we obtain
\be\label{inverse}
\det{|f_1\circ f_2+f_3\circ f_4|}=\ff{2^{4M}}{\det{|f_1\circ
f_4+f_3\circ f_2|}}\,.
\ee
It is convenient to group all terms within the trace into pairs
$(\Phi_1\star \Phi_2)\dots (\Phi_{2n-1}\star \Phi_{2n})$. Taking
the trace, for the determinant one obtains
\be\label{X}
X^2=\prod_{i=1}^{n-1}\ff{1}{\det{|f_{2i-1}+f_{2i}|}}\prod_{i=1}^{n-1}{\ff{\det{|f_1\circ
f_{2i+2}+f_{2i+1}\circ f_{2i}|}}{2^{4M}}}
\ee
Now, property \eqref{cond} peculiar for the propagators states
that the determinant is invariant under simultaneous sign flip
$f_i\to -f_i$ for all $i$. This is equivalent to the statement
that the prefactor in $\textnormal{Tr}(\Phi_1\star \Phi_2)\dots
(\Phi_{2n-1}\star \Phi_{2n})$ is equal to the one taken in the
reverse order $\textnormal{Tr}(\Phi_{2n}\star \Phi_{2n-1})\dots
(\Phi_2\star \Phi_1)$. Equating both expression one arrives at the
following series of identities
\be\label{iden}
\prod_{i=1}^{n-1}{\det{|f_1\circ f_{2i+2}+f_{2i+1}\circ f_{2i}|}}=
\prod_{i=1}^{n-1}{\det{|f_{2n}\circ f_{2i-1}+f_{2i}\circ
f_{2i+1}|}}\,.
\ee
Particularly, for $n=2$ we get
\be
\det{|f_1\circ f_2+f_3\circ f_4|}=\det{|f_2\circ f_1+f_4\circ
f_3|}\,.
\ee
Now, we redefine cyclically all $f_{i}\to f_{i+1}$ and multiply
both expressions \eqref{X} with each other. The result is
\be\label{X2}
\prod\limits_{i=1}^{2n}\det{|f_i+f_{i+1}|}X^{4}=\prod_{i=1}^{n-1}{\ff{\det{|f_1\circ
f_{2i+2}+f_{2i+1}\circ f_{2i}|}}{2^{4M}}}
\prod_{i=1}^{n-1}{\ff{\det{|f_2\circ f_{2i+3}+f_{2i+2}\circ
f_{2i+1}|}}{2^{4M}}}
\ee
Final step is to apply \eqref{iden} and \eqref{inverse} to the
first product in the r.h.s. of \eqref{X2}
\be\label{canc}
\prod\limits_{i=1}^{2n}\det{|f_i+f_{i+1}|}X^{4}=\prod_{i=1}^{n-1}\ff{\det{|f_2\circ
f_{2i+3}+f_{2i+2}\circ f_{2i+1}|}}{\det{|f_{2n}\circ
f_{2i+1}+f_{2i}\circ f_{2i-1}|{2^{4M}}}}
\ee
Note, that the denominator in \eqref{canc} is that of nominator
shifted cyclically by two steps. Therefore both cancel each other
yielding the final result
\be
X_{2n}=\ff{1}{2^{M(n-1)}\sqrt[4]{\prod\limits_{i=1}^{2n}\det{|f_i+f_{i+1}|}}}
\ee
Finally, one has to flip the sign of each $f_{2i}$ in accordance
with $\tilde{\Phi}$ in \eqref{even}.
\end{appendix}

\providecommand{\href}[2]{#2}\begingroup\raggedright\endgroup

\end{document}